\documentclass[12pt]{iopart}

\pdfoutput=1
\usepackage{amsfonts}
\usepackage{graphicx}
\usepackage{hyperref}
\usepackage{color}
\usepackage{subfigure}
\usepackage{textcomp}
\usepackage{cite}
\usepackage{float}
\usepackage{soul}
\usepackage{ulem}
\usepackage{stackrel}
\usepackage{tikz}
\usepackage{lmodern} 
\usepackage{diagbox}
\usepackage{cancel}
\usepackage{bm}
\usepackage{multirow}
\usepackage[mathscr]{eucal}
\usepackage{bbm}

\newcommand{\bcen}{\begin{center}}
	\newcommand{\ecen}{\end{center}}
\newcommand{\btab}{\begin{tabular}}
	\newcommand{\etab}{\end{tabular}}
\newcommand{\bdes}{\begin{description}}
	\newcommand{\edes}{\end{description}}

\newcommand{\beq}{\begin{equation}}
\newcommand{\eeq}{\end{equation}}
\newcommand{\bea}{\begin{eqnarray}}
\newcommand{\eea}{\end{eqnarray}}

\newcommand{\bary}{\begin{array}}
	\newcommand{\eary}{\end{array}}
\newcommand{\benum}{\begin{enumerate}}
	\newcommand{\eenum}{\end{enumerate}}
\newcommand{\bitem}{\begin{itemize}}
	\newcommand{\eitem}{\end{itemize}}

\usepackage{color}
\definecolor{dgreen}{rgb}{0,0.7,0}
\definecolor{dc}{rgb}{0.8,0.1,0.3}
\def\redw#1{{\color{red} #1}}

\expandafter\let\csname equation*\endcsname\relax
\expandafter\let\csname endequation*\endcsname\relax
\usepackage{amsmath,amssymb}

\pdfoutput=1
\usepackage{esint}

\newcommand{\titlename}{Crossover in densities of confined particles with finite range of interaction}

\begin{document}
	
	\title{\titlename}
	
	\author{Saikat Santra and Anupam Kundu}

	\address{International Centre for Theoretical Sciences, Tata Institute of Fundamental Research, Bengaluru -- 560089, India}

 \ead{saikat.santra@icts.res.in, anupam.kundu@icts.res.in} 
	
	

\date{\today}
	
\begin{abstract}
We study a one-dimensional classical system of $N$ particles  confined within a harmonic trap. Interactions among these particles are dictated by a pairwise potential $V(x)$, where $x$ is the separation between two particles. Each particle can interact with a maximum of $d$ neighbouring particles on either side (left or right), if available. By adjusting the parameter $d$, the system can be made nearest neighbour $(d=1)$ to all-to-all $(d=N-1)$ interacting. As suggested by prior studies, the equilibrium density profile of these particles is expected to undergo shape variations as $d$ is changed. In this paper, we investigate this crossover by tuning the parameter $f(=d/N)$ from $1$ to $0$ in the large $N$ limit for two distinct choices of interaction potentials, $V(x) = - |x|$ and  $V(x) =- \log(|x|)$ which correspond to 1d one-component plasma and the log-gas model, respectively.  For both models, the system size scaling of the density profile for fixed $f$ turns out to be the same as in their respective all-to-all case. However, the scaling function exhibits diverse shapes as $f$ varies. We explicitly compute the average density profile for any $f \in (0,1]$ in the 1d plasma model, while for the log-gas model, we provide approximate calculations for large (close to $1$) and small  (close to $0$) $f$. Additionally, we present simulation results to numerically demonstrate the crossover and compare these findings with our theoretical results.

\end{abstract}	
\maketitle
\section{Introduction}
\label{Introduction}
The study of dynamics and thermodynamic properties of long-range interacting systems had generated lots of interests in the last few decades \cite{Cohen_PRE_2001,Ruffo_springer_2002,Campa_physica_2004,Lapo_physica_2007,Lapo_JSM_2010,Marcos_PRL_2010,Latella_PRE_2013}. Often these systems are characterized by the two-particle interaction potential $V(x)\sim |x|^{-k}$
where $x$ is the distance between two particles. The parameter $k$ determines the range of interaction --- smaller the value of $k$,  larger the range of the interaction potential. 
One of the well known example of many particle classical interacting system in one dimension is harmonically confined  Riesz gas consisting of $N$ particles and it is described by the energy function \cite{riesz_ASU_1938}
\begin{align}
\mathscr{E}(\{x_j\})= \frac{1}{2} \sum_{i=1}^{N} x^2_i+ \frac{J ~ \rm {sgn}(k)}{2} \sum_{\substack{i\neq j=1\\|i-j| \leq d}}^{N} \frac{1}{\mid x_i-x_j \mid ^k}, ~\text{with}~k>-2, \label{gen_model} 
\end{align}
where ${x_i}$ with $i=1,2,...,N$ denote the positions of the particles, $J>0$ controls the strength of the interaction. The term ${\rm sgn}(k)$ ensures a repulsive interaction. The parameter $d$ in Eq.~\eqref{gen_model} determines  the number of particles each particle is allowed to interact with on either side (left or right) of it, if available.  For $d \sim \mathcal{O}(1)$, each particle can interact only with a few other particles. Such a system is called short ranged (SR). When $d=N-1$, the energy function in Eq.~\eqref{gen_model} reduces to the all-to-all coupled (ATAC) model  in which each particle interacts with every other particles present in the system. By tuning the fraction $f=\frac{d}{N}$ from $0$ to $1$ in the large $N$ limit, one can go from the SR regime $(f=0)$ to the  ATAC regime ($f=1$). We call systems with intermediate values of $f$  {\it i.e.,} $0<f<1$ to be finite ranged (FR).

Special values of $k$ in the ATAC case ($f = 1$) correspond to important models which has been studied both in physics and mathematics.  For example, taking the limits $k\to0, J  \to \infty$ such that $J|k| \to J_0~(>0)$ in Eq.~\eqref{gen_model} with $d=N-1$ one obtains a logarithmic interaction potential which corresponds to the well known log-gas model \cite{dyson_JMP_1962,dyson_JKP_1962,mehta2004random,forrester_PUP_2010}. This system has a direct connection to the random matrix theory~\cite{mehta2004random,forrester_PUP_2010}and has been studied in diverse contexts, such as, extreme value statistics \cite{Dean_PRL_2006,Dean_PRE_2008,Majumdar_IOP_2014}, index distribution~\cite{Majumdar_PRL_2009}, number variance~\cite{Phase_PRL_marino}, third-order phase transition ~\cite{Majumdar_IOP_2014,Cunden_JSM_2017,Cunden_Springer_2019}, non-interacting fermions \cite{Dean_PRA_2016,Dean_EPL_2019,Dean_JPA_2019}, communication system~\cite{Pavlos_IEEE_2011}, large-$N$ guage theory \cite{Douglas_nuclear_1995} to name a few. 

Riesz gas with $k=2$ and $d=N-1$ corresponds to the  Calogero Moser system which is a well known model for integrable  interaction potential~\cite{Calogero_JMP_1969,Calogero_JMP_1971,Calogero_LNC_1975, polychronakos_IOP_2006}.
The value $k=-1$ in the ATAC case corresponds to the one-dimensional one-component plasma ($1$dOCP), also known as the Jellium model~\cite{Lenard_JMP_1961,baxter_phil_1963}, in which the interaction energy between two particles decreases linearly with increasing separation between them. The $1$dOCP model has been studied previously to understand statistical properties of plasma \cite{Lenard_JMP_1961,baxter_phil_1963}. In this model, researchers have recently explored extreme value statistics and index distribution~\cite{dhar_IOP_2018, dhar_PRL_2017}, truncated linear statistics~\cite{Flack_JPA_2021}, full counting statistics~\cite{Flack_JSM_2022}, as well as demonstrated a third-order phase transition~\cite{dhar_PRL_2017,Cunden_JPA_2018}. 
		
\noindent In thermal equilibrium  at temperature $T=\beta^{-1}$, the Riesz gas system  in Eq.~\eqref{gen_model} can be described by the Boltzmann distribution 
\beq
P_G(\{x_j\}) =e^{-\beta \mathscr{E}(\{x_j\})}/Z_N(\beta)
\label{eq_boltzmann},
\eeq
where $Z_N(\beta)$ is the partition function. We assume Boltzmann constant $k_B = 1$ in this paper. 
The first natural question: what is the equilibrium density profile of the particles? For the ATAC case ($f=1$), by developing a large-$N$ field theory,  the average thermal density has been calculated for any $k>-2$ in Ref.~\cite{sanaa2019} at small temperature [$\beta^{-1} \sim \mathcal{O}(1)$]. It was observed that the average thermal density for large $N$ is independent of the temperature of the system and has a finite support for all $k>-2$, although the support explicitly depends on the power-law exponent $k$~\cite{sanaa2019}.  Several other studies have also been conducted on Riesz gas in the ATAC case. These include particle fluctuations in log-gas~\cite{GUSTAVSSON_2005}, full particle statistics in $1$d Coulomb gases with an arbitrary external potential~\cite{Rojas_PRE_2018}, density profile in presence of a hard wall~\cite{jitendra_JSM_2021}, extreme value statistics~\cite{jitendra_JSM_2022}, universality in third-order phase transition~\cite{jitendra_JSM_2022} and gap statistics~\cite{saikat_PRL_2022}. Experimental realizations of systems with some fractional values of $k$ have also been achieved~\cite{Joseph_PRL_2011,zhang_nature_2017}.

While these results are valid for ATAC case of Riesz gas {\it i.e.,} for $f=1$, they are expected to get modified for other values of $f <1$. Moreover, in most of the physical systems the particles do not interact with all other particles and hence the interactions are not all-to-all coupled. It is important to study the effect of the parameter $f$ on different physical quantities of the system. The basic physical quantity that one would naturally consider is the average equilibrium density profile and study how the density profile changes with $f$ in the large $N$ limit. 
Recently, the equilibrium density profile for the SR case ($f=0$) has been computed in Ref.~\cite{Avanish_PRE_2020,Pandey_PRE_2020}. In this paper, after developing a large-$N$ field theory similar to the all-to-all coupling case, the density profile was obtained analytically for $\beta \sim \mathcal{O}(1)$. It was found that in the SR case the density profile is drastically different  compared to the ATAC scenario for Riesz gas with $k \le 0$. However, interestingly, for $k>0$ the shape of the density profile in the SR case remains same as in the ATAC case~\cite{Avanish_PRE_2020}. Natural questions arise: what is the density profile for intermediate $f$ {\it i.e.,} for $0<f<1$? How does the shape of density profile change  as $f$ is tuned from $1$ to $0$ while keeping the temperature fixed at $\mathcal{O}(1)$ value? In this paper, we address this question for two specific systems: $1$dOCP and log-gas which correspond to $k=-1$ and $k \to 0$ limit of the energy function in Eq.~\eqref{gen_model}, respectively. A similar question was studied in the context of random matrix theory in Ref.~\cite{Invariant_PRL_Allez}, where a change in the spectral density profile from the Wigner semi-circle to a Gaussian form was observed as the strength of the interaction was varied smoothly from a very high value to a low value. With decreasing interaction strength, the entropy starts taking  part in balancing the confining effect of the harmonic `trap'.  At very small value of the interaction strength, only entropic contribution is dominant and one observes a Gaussian distribution as one would see for non-interacting particles inside a harmonic trap. Same  interaction strength (or equivalently temperature) tuned crossover of spectral density was observed in the context of invariant $\beta$-Wishart ensembles as well~\cite{Allez2013}. 

For  the $1$dOCP and log-gas models, we demonstrate a crossover in the density from a finitely supported profile to a Gaussian form as the  parameter $f$ is tuned from $1$ to $0$. For the $1$dOCP model we obtain analytical expression for the density profile for any $f \in (0,1]$ in the large $N$ limit, while for the log-gas model we  provide approximate analytical calculations to understand the density profiles for $f$ close to $1$ and $0$, separately. For other values of $k>-2$, we numerically  demonstrate a similar crossover in the density profile. 		
			
The rest of the paper is organized 	as follows. In Sec.~\ref{model_and_definitions}, we define the main physical quantity of interest  and give a quick summary of important previous results. We provide the summary of our findings in Sec.~\ref{quantity_summary} along with numerical demonstrations of the density crossover in both systems.   In Sec.~\ref{density_crossover}, we study the density crossover in $1$dOCP ($k=-1$)  by solving saddle point equations for $f \in (0,1]$. We present this solution in two parts, for $1/2 \leq f \leq 1$ in Sec.~\ref{saddle-1docp-text-f>1/2} and for $0<f<1/2$ in Sec.~\ref{saddle-1docp-text-f<1/2}, respectively. Then in Sec.~\ref{crossover_genkgt0}, we study the density crossover in log-gas ($k\to0$) model. Along with the numerical results, we provide approximate analytical calculations to understand the density profiles both in the  $f\to1$ and $f \to 0$ limits. Finally in Sec.~\ref{conclusions}, we conclude our study along with some interesting future directions.  Some details of the calculations and numerical simulations are relegated to the Appendices.  
The details of numerical methods used in this paper are given in \ref{numerical_details}. The derivation of the saddle point equations  are presented in \ref{sadle-eq-derv}.
Some details pertaining to the solution of the saddle point equations in the $1$dOCP case are provided in \ref{Solution_sp_1docp}. In \ref{app:den-cross-other-k}, we numerically demonstrate similar density crossover in the general  Riesz gas model defined in Eq.~\eqref{gen_model} for other values of $k$.		
	
\section{Definitions and relevant previous results} 
\label{model_and_definitions}
We consider FR Riesz gas consisting of $N$ particles in thermal equilibrium at temperature $\beta^{-1} \sim \mathcal{O}(1)$. We focus on the two choices of the interaction potentials, $1$dOCP and log-gas for which the microscopic energy functions are given explicitly as 
\begin{align}
\begin{split}
\mathscr{E}(\{x_j\})&= \frac{1}{2} \sum_{i=1}^{N} x^2_i+ \frac{1}{2} \underset{j\ne i}{\sum_{i=1}^N \sum_{j=\max(i-d,1)}^{j=\min(i+d,N)}}~V(x_i-x_j), \\
~\text{with}~&~
V(x) =
\begin{cases}
  ~~-|x|, ~&~\text{for $1$dOCP}, \\
-\log(|x|)~&~\text{for log-gas}.
\end{cases}
\end{split}
\label{E-ocp-lg} 
\end{align}
In Eq.~\eqref{E-ocp-lg},
we have set the strengths $J$ and $J_0$ of the interaction potentials, respectively, for $1$dOCP and log-gas to $1$ in this paper.

In these systems, we are interested to compute the average equilibrium density profile $\varrho_N(x) = \langle \rho_N(x)\rangle$, where the empirical density $ \rho_N(x)$ is defined as $\rho_N(x) \equiv \frac{1}{N}\sum_{i=1}^{N} \delta(x-x_i)$, and the average $\langle...\rangle$ is performed over the Boltzmann distribution in Eq.~\eqref{eq_boltzmann}. In the large $N$ limit, the multiple integrals required in order to compute the average $\langle ... \rangle$, can be converted to a functional integral over fluctuating density profiles weighted appropriately by a free energy functional $\Sigma[\rho_N(x)]$ written in terms of the empirical density, 
 $\rho_N(x)$ \cite{Dean_PRE_2008,sanaa2019, Avanish_PRE_2020}. The free energy functional $\Sigma [\rho_N(x)]$ for a given density profile $\rho_N(x)$ has two parts 
  \begin{align}
\Sigma [\rho_N(x)]=\mathcal{E}[\rho_N(x)] - \beta^{-1} \mathcal{S}[\rho_N(x)],   \label{eq:Sigma}  
 \end{align} 
where $\mathcal{E}[\rho_N(x)]$ is the energy functional and  $\mathcal{S}[\rho_N(x)] = -N\int_{-\infty}^{\infty}  dx \rho_N(x) \log[\rho_N(x)]$  is the entropy functional~\cite{Dean_PRE_2008}. The energy functional needs to be determined from the microscopic energy function in Eq.~\eqref{E-ocp-lg} and in general should contain a self energy term and  the bulk energy term~\cite{Sandier_arxiv_2013,Dean_PRE_2008}.

In terms of the free energy functional $\Sigma [\rho_N(x)]$, the partition function of the system can be written (up to an overall multiplicative factor) as~\cite{Dean_PRE_2008, sanaa2019, Avanish_PRE_2020}
	\beq
	Z_N(\beta) \sim \int d\mu \int \mathcal{D}[\rho_N(x)]	\exp\Big [-\beta \bar{\Sigma}[\rho_N(x)] \Big],
	\eeq where $ \int \mathcal{D}[\rho_N(x)]$ denotes functional integration over density function $\rho_N(x)$ and 
 \begin{align}
\bar{\Sigma}[\rho_N(x)]=  \Sigma[\rho_N(x)]-\mu\left( \int dx \rho_N(x) -1\right).    \label{Sigma_mu} 
 \end{align} 
 Note $\mu$ is a Lagrange multiplier  that ensures the normalisation of the density profile to unity.  In the large $N$ limit, the equilibrium density profile $\varrho_N(x)$ can be obtained by solving the saddle point equations (SPEs) \cite{sanaa2019,Avanish_PRE_2020} \begin{align}
 \frac{\delta \bar{\Sigma}[\rho_N(x)]}{\delta \rho_N(x)}\big{|}_{\rho_N(x)=\varrho_N(x)}=0,~~\text{and}~~\frac{\partial \bar{\Sigma}}{\partial \mu}=0.
 \label{spe}
 \end{align}
For fixed $f$ it is possible to show that in the large $N$ limit, the energy functional  scales as $\sim N^{\lambda}$ with $\lambda \ge 1$. This fact has been proved for Riesz gas with general $k (>-2)$ for $f = 1$ and $f=0$ in Ref.~\cite{sanaa2019} and Ref.~\cite{Avanish_PRE_2020}, respectively. Additionally, in the large $N$ limit one can show (see \ref{sadle-eq-derv}) that for fixed $f \in (0,1]$ the contributions from the entropy and the self energy are much smaller compared to the bulk energy term and consequently, they can be neglected while solving the SPEs.

Before going into the details of the derivation and presentation of the results for the FR case ($0<f<1$), we discuss some of the relevant results obtained previously for ATAC ($f=1$)  and SR ($f = 0$) cases.
 	
\subsection{Equilibrium density profile in the ATAC case ($d=N-1$ {\it i.e.,} $f=1$)}
 \label{a2a-previous}
In this case, the entropy as well as the self energy terms are subdominant compared to the bulk energy part in the free energy \cite{Dean_PRE_2008, sanaa2019}. Neglecting these contributions, minimizing the free energy functional essentially becomes minimizing the (bulk) energy functional. One finds that the average density $ \varrho_N(x)$ is described, for both $1$dOCP and log-gas systems, by the following scaling form~\cite{sanaa2019}
\beq
\varrho_N(x)  = \frac{1}{ N^{\alpha}} \tilde{\varrho}_{1} \left(\frac{x}{N^{\alpha}}\right),~~\text{with}~~
\alpha  =	\begin{cases} 1 ~~~~{\rm for}~~~~\text{$1$dOCP} \\
\frac{1}{2} ~~~~{\rm for}~~~~\text{log-gas},
\end{cases}	
\label{density_all_to_all}
\eeq
with scaling functions  given explicitly by
\begin{align}
\tilde{\varrho}_1(y)= \begin{cases} ~~~~~~~\frac{1}{2}~~~~~~~~~~~{\rm for}~~~~-1 \le y \le 1~~~~~~~\text{in $1$dOCP}\\
~\frac{1}{\pi} \sqrt{2-y^2} ~~~~~~{\rm for}~~-\sqrt{2} \le y \le \sqrt{2}~~~~\text{in log-gas}. 
\end{cases}
\label{sc-den-f-1}		
\end{align}
We would like to emphasize that these scaling functions correspond to the equilibrium densities of particles under conditions of $\mathcal{O}(1)$ temperature and large $N$. In such scenarios, the contribution from entropy can be disregarded when solving the SPEs. However, as the temperature increases, the entropy's influence becomes increasingly significant. Consequently, the solution of the SPE deviates from those presented in Eqs.~\ref{density_all_to_all} and \ref{sc-den-f-1}. This phenomenon has been rigorously examined within the framework of random matrix theory \cite{Invariant_PRL_Allez, Allez2013}, wherein instead of raising the temperature, the interaction strength is diminished. This adjustment leads to a cooperative interplay between energy and entropy, working together to counterbalance the confinement imposed by harmonic trapping.

\subsection{Equilibrium density profile in SR case ($d\sim \mathcal{O}(1)$ {\it i.e.,} $f=0$)} 
\label{sr-previous}
\noindent Contrary to the previous case, the entropy term in this case dominates over the interaction energy term  in the $1$dOCP system. However for the log-gas the entropy, the self energy and the bulk interaction energy all contribute at the same order \cite{Avanish_PRE_2020}. The density profiles in this case for large-$N$ are given by  \cite{Avanish_PRE_2020}
\begin{align}
\varrho_N (x) =\varrho_0(x),~~\text{where},~~\varrho_0(x)= \begin{cases}
~~~~~\sqrt{\frac{\beta}{2 \pi}} \text{exp}(-\beta x^2/2) ~~ &\text{\rm for $1$dOCP}  \\ 
\sqrt{\frac{\beta}{2 (\beta d+1) \pi}} \text{exp}(-\frac{\beta x^2}{2(\beta d+1)} )
 ~~ &\text{\rm for log-gas}. \end{cases}
\label{eq:denk<0}
\end{align}
In contrast to the ATAC scenario ($f=1$), the equilibrium density profiles in the SR case remain independent of the system size $N$ and are of Gaussian form.

\begin{figure}[t]
	\centering		
	\includegraphics[width=\textwidth]{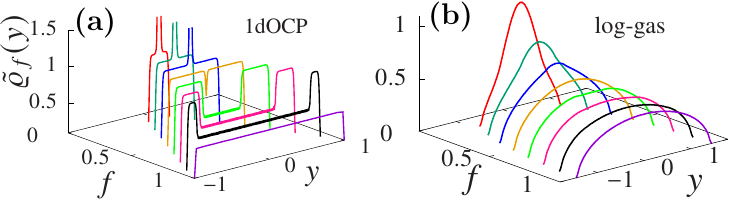}
	\caption{Demonstration of the crossover in the density profiles in (a) $1$dOCP  and (b) log-gas systems. For both the systems we plot the scaled density profile $\tilde{\varrho}_f(y)=N^\alpha \varrho_N(yN^\alpha)$ as functions of $y$ [see Eq.~\eqref{den-sc-form-fr}] for eight values of $f=n/8$ with $n=1$ (red), $2$ (dark-green), $3$ (blue), $4$ (brown), $5$ (green), $6$ (dark-pink),  $7$ (black), and $8$ (violet). For $1$dOCP model, the profile transitions from a flat profile at $f=1$ to a delta function profile in the $f \to 0$ limit.
The peaks of the delta functions on the $z$-axis are cut in order to display the features of the density profiles for $f$ close to $1$. For the  log-gas, the  density profile changes its shape [shown in figure (b)] from Wigner semi-circle at $f=1$ to bell-shaped at small $f$.  For both cases we observe that the total support of the density profile decreases as $f$ is reduced. The plots of density profiles are obtained numerically for $N=128$ for both the systems. Details of the numerical simulation are provided in \ref{numerical_details}.}		
	\label{density_summary}
\end{figure}

\section{Summary of the results}
\label{quantity_summary}
As mentioned in the previous section, the equilibrium density profiles are dramatically different for ATAC ($f=1$) and the SR ($f = 0$) case. In this paper, we investigate how the density profiles cross over from  finitely supported profiles to  infinitely extended Gaussian profiles as $f$ is decreased from $1$ to $0$. We find that in both models, the density profiles  for large $N$ and  fixed $f$ consist of three parts: one central part extending over region $x \in [-\ell_N,\ell_N]$ and two symmetric edge parts supported over regions $[-b_N,-\ell_N)$ and $(\ell_N,b_N]$, respectively, where $0<\ell_N<b_N$. At low temperature ($T \sim \mathcal{O}(1)$), these three regimes naturally appear because interaction felt by a particle in these regimes are different. For $f<1$, the central part contains $|1-2f|$ fractions of particles where each particle can interact 
either with $2d$ other particles or the remaining $(N-1)$ particles depending on whether $2d<(N-1)$ or not. 
On the other hand, the remaining fraction of particles residing on the edges always find less than $d$ particles on at least one side, either left or right. 
The lengths of these supports depend on $f$. As $f$ is changed, the ranges as well as the shape of the profiles in these regimes get modified and one observes the above mentioned crossover in the density profile
as numerically demonstrated in Fig.~\ref{density_summary} for both $1$dOCP and log-gas models. Below, we briefly summarise our main findings.

\begin{enumerate}
\item We find that for both $1$dOCP and log-gas, the equilibrium density profile for  fixed $f$ in the large $N$ limit, possesses the following scaling form,
\beq
\varrho_N(x) = \frac{1}{N^\alpha}~\tilde{\varrho}_f\left(\frac{x}{N^\alpha}\right),~~\text{with}~
\begin{cases}
\alpha = 1,~&~\text{for $1$dOCP}, \\
\alpha=\frac{1}{2},~&~\text{for log-gas}.
\end{cases}
\label{den-sc-form-fr}
\eeq
Note even for $f<1$, the density profile satisfies the same scaling form as in the ATAC case given in Eq.~\eqref{density_all_to_all}. The length of the total support of the scaling function $\tilde{\varrho}_f(y)$ remains finite for all $f>0$ and it decreases  as $f$ is reduced from $1$ (see Fig.~\ref{density_summary}).

\item We have derived the saddle point equations satisfied by the equilibrium density profiles for both cases (see ~\ref{sadle-eq-derv}). 

\item For the $1$dOCP case, one can solve the SPEs for arbitrary values of $f \in (0,1]$. We find that the scaling function 
$\tilde{\varrho}_f(y)$, in this case,  is given by 
\begin{eqnarray}
     \tilde{\varrho}_f(y)&=\begin{cases}
     1 &\text{for}~ -f  \le y<-(2f-1), \\
     \frac{1}{2} & \text{for}~ -(2f-1) \le y \le (2f-1), \\
     1 & \text{for}~ ~~~(2f-1)<y \le f,
 \end{cases}
 ~&~\text{for}~~\frac{1}{2}\leq f \leq 1,~~~ \label{trho-k_-1-final-f>1/2}\\ 
 & & \notag \\
 \tilde{\varrho}_f(y)&=(1-2f) \delta(y)+ \Theta(y+f)\Theta(f-y),~&~\text{for}~~0<f < \frac{1}{2}, ~~~\label{trho-k_-1-final-f<1/2}
\end{eqnarray}
which describes the crossover  in the scaled density profile as $f$ is reduced from $1$ to $0$. This crossover phenomena is demonstrated in Fig.~\ref{density_summary}(a).  In the $f\to0$ limit, the strength of the delta function part grows and approaches unity which (as shown in Sec.~\ref{saddle-1docp-text-f<1/2}),  when zoomed in,  essentially corresponds to the Gaussian profile given in Eq.~\eqref{eq:denk<0} in terms of the unscaled variable $x$  as obtained previously in Ref.~\cite{Avanish_PRE_2020}. 

\item For the log-gas case, we numerically demonstrate the crossover of density profile from Wigner semi-circle to Gaussian form as $f$ is decreased from $1$ to $0$ (see Fig.~\ref{density_summary}(b)). Solving the SPEs for arbitrary values of $f$ seems difficult in this case. We find  approximate solutions for the density profiles for $f$ close to $1$ and $0$.  
In the former case, we find that the scaling function for the density can approximately be described  by~\cite{Phase_PRL_marino}
\beq
\tilde{\varrho}_f(y) \approx \frac{1}{\pi} \sqrt{\frac{(y^2-\tilde{a}^2)(\tilde{b}^2-y^2)}{(y^2-\tilde{\ell}^2)}},
\label{density_index_satya}
\eeq
which exists on a three-cut support $[-\tilde{b},-\tilde{a}],~[-\tilde{\ell},\tilde{\ell}],$ and $[\tilde{a},\tilde{b}]$  with $0<\tilde{\ell}<\tilde{a}<\tilde{b}$ and 
$y$ belongs to any of the intervals in the support. For given $\tilde{\ell}$, the parameters $\tilde{a},\tilde{b}$ can be determined from the relation $\tilde{a}^2+\tilde{b}^2=2+\tilde{\ell}^2$ and the normalisation condition $\int_{\tilde{a}}^{\tilde{b}}\tilde{\varrho}_f(y)dy=(1-f)$~\cite{Phase_PRL_marino}. In our case we determine $\tilde{\ell}$ from the numerically obtained density, $\tilde{\varrho}_{\rm nu}(y)$ using the relation $\int_{-\tilde{\ell}_{\rm nu}}^{\tilde{\ell}_{\rm nu}}\tilde{\varrho}_{\rm nu}(y) dy=(2f-1)$,
where the subscript `nu' stands for numerical value. With increasing $f$ towards unity, the parameters  $\tilde{\ell}$ and $\tilde{a}$ in Eq.~\eqref{density_index_satya} approach to $\tilde{b}$. Exactly at $f=1$, they all become equal to $\sqrt{2}$ and one recovers the density profile in the ATAC case (given in Eq.~\eqref{sc-den-f-1}). A comparison of our theoretical result with numerically obtained density  is provided in Fig.~\ref{largef_kzero_figure} where we observe descent agreement.

\hspace{0.3cm}
For small $f$, we find that the density profile in the central part can be well described by the following approximate scaling function  
\beq
\tilde{\varrho}_f(y) \approx 
 \frac{1}{\sqrt{2\pi f}}\exp\left(-\frac{y^2}{2 f}\right),~~-\tilde{\ell} \leq y \leq  \tilde{\ell},
\label{smallf_kzero_theory}
\eeq
where $\tilde{\ell}$ is determined from $ \int_{-\tilde{\ell}}^{\tilde{\ell}} \tilde{\varrho}_f(y)dy =1-2f$. This result is consistent with the result in Eq.~\eqref{eq:denk<0} obtained previously in Ref.~\cite{Avanish_PRE_2020}. To see this one needs to consider both $d$ and $N$ large in Eq.~\eqref{eq:denk<0}, so that the fraction $f=d/N$ is fixed. This essentially  means $\tilde{\ell} \to \infty$ as $f \to 0$  in Eq.~\eqref{smallf_kzero_theory}. To determine $\tilde{\varrho}_f(y)$ outside the central part, we notice that  for small $f$ the solution of the SPE can be approximated by a problem in ATAC  log-gas system with a region $[-\tilde{\ell},\tilde{\ell}]$ deprived of particles.  The saddle point solution for this problem is given by~\cite{Phase_PRL_marino}
\beq
\tilde{\varrho}_f(y) \approx  \frac{1}{\pi} \sqrt{\frac{y^2~(\tilde{b}^2-y^2)}{(y^2-\tilde{\ell}^2)}} \text{~~for~~} \tilde{\ell} < |y| \leq \tilde{b},
\label{density_edge_log}
\eeq
where $\tilde{b}=\sqrt{\tilde{\ell}^2+4f}$ can be obtained from the constraint $ \int_{\tilde{\ell}}^{\tilde{b}} \tilde{\varrho}_f(y)dy =f$. We find that this form  describes the behaviour of the density profile near the edges well (see Fig.~\ref{smallf_kzero_figure}). 
	
\end{enumerate}

\section{Density crossover in $1$dOCP ($V(x)=-|x|$)}
\label{density_crossover}
In this section, we solve the SPE in Eq.~\eqref{spe} for given $f$ to study crossover in the density profile for the $1$dOCP model described by the  energy function given in Eq.~\eqref{E-ocp-lg}. For $0<f<1$ one gets separate SPEs in the three parts:  the central part $[-\ell_N,\ell_N]$, the left edge part $[-b_N,-\ell_N)$ and the right edge part $(\ell_N,b_N]$. These equations, derived in \ref{sadle-eq-derv}, can be solved exactly as shown below.
We first present the solution for $1/2 \leq f \leq 1$. For $0<f<1/2$, as we will see, one requires to solve the SPEs separately for  $1/3 \leq f<1/2$ and $0<f<1/3$.

\subsection{Solution for $1/2\leq f \leq1$:}
\label{saddle-1docp-text-f>1/2}	
\begin{figure}[t]
	\centering		
	\includegraphics[width=1.0\textwidth]{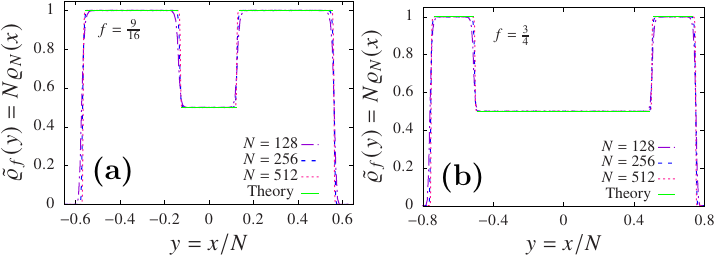}		
	\caption{Plots of numerically obtained density profiles (dashed lines) for (a) $f=\frac{9}{16}$ and  (b) $f=\frac{3}{4}$ in $1$dOCP. Collapse of data for different $N$ verifies the scaling relation in Eq.~\eqref{den-sc-form-fr} with $\alpha=1$.  The solid green lines represent the analytical prediction in Eq.~\eqref{trho-k_-1-final-f>1/2}. We observe good agreement between simulation and theory. With decreasing $f$, the size of the central region reduces and the edge regions grow.} 
	\label{largef_kmone_figure}
\end{figure}

For fixed $f$ and large $N$, in order to minimise energy (see  Eq.~\eqref{E-ocp-lg}), particles settle over a region which scales as $\mathcal{O}(N)$. In this configuration, the contributions from the confining harmonic term and the repulsive interaction term both become order $N^3$ such that they can compete each other. Note, on the other hand entropy is always of order $N$ for $\beta^{-1} \sim \mathcal{O}(1)$. Hence it is expected that the equilibrium density profile possesses the scaling $\varrho_N(x) = \frac{1}{N} \tilde{\varrho}_f(x/N)$ which is also verified numerically in Fig.~\ref{largef_kmone_figure} for two representative values of~$f>1/2$. To find the scaling function $\tilde{\varrho}_f(y)$, it seems convenient to rewrite the SPEs in terms of the scaling variable $y=x/N$ and the scaling functions $\tilde{\varrho}_f(y)$ as given in  Eqs.~\ref{sp_1docp_flarge_central_scaled2} and \ref{eq:sp_1docp_flarge_edge_scaled}. Furthermore, numerical observations from Fig.~\ref{largef_kmone_figure} suggests that the equilibrium density profile is piece-wise uniform. This leads us to make the following ansatz for the saddle point density profile 
\beq
 \tilde{\varrho}_f(y)=\begin{cases}
     \tilde{\varrho}_{\text{edge}} ~\text{for}~ -\tilde{b} \le y<-\tilde{\ell} \\
     \tilde{\varrho}_{\text{mid}} ~~\text{for}~ -\tilde{\ell}\le y \le \tilde{\ell} \\
     \tilde{\varrho}_{\text{edge}} ~~\text{for}~ ~~~\tilde{\ell}<y \le \tilde{b},
 \end{cases}
 \label{density_piecewise_scaled}
 \eeq
where $\tilde{\ell}=\ell_N/N$ and $\tilde{b}=b_N/N$. Note there are six unknown constants, namely, $\tilde{\varrho}_{\text{edge}},~\tilde{\varrho}_{\text{mid}},~\tilde{b},~\tilde{\ell}$ and the chemical potentials $\tilde{\mu}_1$ and $\tilde{\mu}_2$, which are required to ensure appropriate normalisations of the density profiles in respective parts [see Eq.~\eqref{constraint_fgt1}]. Inserting the above form of the scaled density profile in SPEs, one finds six equations which determine the values of these constants. Since the calculations to find these six constants are involved and lengthy, we present the details of  this calculation  in \ref{sp-sol-fr1docp-f>1/2}. Here, we only present the final result:
\begin{align}
\begin{split}
&\tilde{\varrho}_{\text{edge}} = 1, ~~~~~~
\tilde{\varrho}_{\text{mid}} = \frac{1}{2},  \\
&\tilde{\ell}=(2f-1), ~~~
\tilde{b}=f, \\
&\tilde{\mu}_1 = \tilde{\mu}_2 = -2 \left(f -\frac{1}{2}\right)^2,
\end{split}
\label{sol-sd-eq-f>1/2}
\end{align}
inserting which in Eq.~\eqref{density_piecewise_scaled} completely determines $\tilde{\varrho}_f(y)$ for $1/2\leq f \leq 1$. The theoretical density profile is numerically verified in Fig.~\eqref{largef_kmone_figure}.

\subsection{Solution for $0<f<1/2$:} 
\label{saddle-1docp-text-f<1/2}
\begin{figure}[t]
	\centering		
	\includegraphics[width=1.0\textwidth]{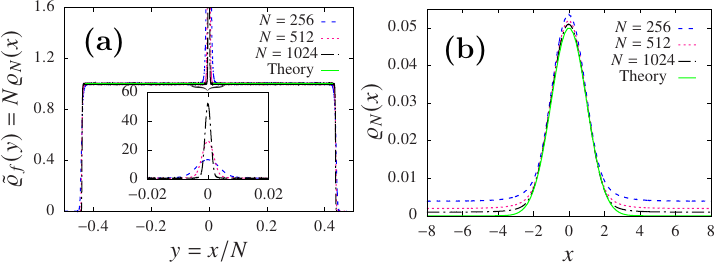}		
	\caption{(a) System size collapse of scaled density $\tilde{\varrho}_f(y)=N \varrho_N(x)$ for $1$dOCP at $f=7/16$. Data collapse for different $N$  verifies the scaling relation in Eq.~\eqref{den-sc-form-fr} with $\alpha=1$ over almost the full range except for a small region around the origin. The green solid line represents a uniform profile  on both the left and right edge parts, as described by Eq.~\eqref{trho-k_-1-final-f<1/2}. The inset displays a zoomed version of the central part, where we observe that under scaling $y=x/N$ the density profile for larger $N$ becomes sharper and narrower which is in consistent with the delta function in Eq.~\eqref{trho-k_-1-final-f<1/2}. (b) This figure plots the density profile in the  central part [zoomed version of delta part in (a)] as functions of unscaled variable $x$ for different $N$. We observe that the density profile in the central part approaches, with increasing $N$, to a Gaussian form as given in Eq.~\eqref{density_kmone_fless}. Simulation data are obtained using  $\beta=1$.
	}
	\label{smallf_kmone_figure}
\end{figure}

In this case also we find that the equilibrium density profile has scaling form $\varrho_N(x) = \frac{1}{N} \tilde{\varrho}_f(x/N)$ which is  demonstrated numerically in Fig.~\ref{smallf_kmone_figure}(a). Note that with increasing $N$, the central part (in terms of the scaling variable $y=x/N$) shrinks to a delta function (see inset of Fig.~\ref{smallf_kmone_figure}(a)) while the profiles in the edge parts are uniform. When this delta function is zoomed to the original scale, we observe from Fig.~\ref{smallf_kmone_figure}(b) that, the density profile there can  very nicely be described by a Gaussian form in terms of the unscaled variable $x$. This suggests us to expect the following form for the equilibrium density profile (in terms of the unscaled variable $x$)
\begin{align}
&\varrho_N(x) = 
\begin{cases}
\rho_{\rm central}(x) &~\text{for}~-\ell_N \le x \le \ell_N \\
\rho_{\rm edge} &~\text{for}~~~~\ell_N < |x| \le  b_N,
\end{cases}
\label{rho(x)-ansatz)}
\end{align}
such that $\int_{-\ell_N}^{\ell_N} dx\rho_{\rm central}(x)=1-2f$, $\rho_{\rm edge}=f/(b_N-\ell_N)$. 
This, in turn, suggests us the following form for the scaled density profile: 
\beq
\tilde{\varrho}_f(y)=(1-2f) \delta(y)+\tilde{\varrho}_{\text{edge}} \mathbbm{1}_{[-\tilde{b},\tilde{b}]}(y),
\label{eq:density_k_mone_scaled_text}
\eeq
where $\tilde{\ell}=\ell_N/N \to 0$ and, $\tilde{\varrho}_{\rm edge}=N \rho_{\rm edge}$ and $\tilde{b} = b_N/N$. Note, $\tilde{\varrho}_{\rm edge}$ and $\tilde{b}$ are  $\mathcal{O}(1)$ constants which we need to find. We also show below that $\rho_{\rm central}(x)$ indeed has a Gaussian form. We first write the SPEs from Eqs.~\eqref{SPE_genk_f<1/2_left}-\eqref{SPE_genk_f<1/2_center} (with $V(x)=-|x|$) in terms of the scaling variable $y=x/N$ and the scaled density profile $\tilde{\varrho}_f(y)$. The explicit expression of the SPE on the left edge part $[-\tilde{b},-\tilde{\ell})$ is given in Eq.~\eqref{eq:se_fless_edge_kmone_scaled}. Here, we rewrite it for convenience:
\begin{align}
&\frac{y^2}{2}-  \int_{-\tilde{b}}^{0_-} |y-y'|  \tilde{\varrho}_f(y') dy' \notag \\
& - \Big[\int_{0_-}^{y+\bar{\delta}_y} (y'-y) \tilde{\varrho}_f(y') dy'
 -\frac{1}{2}\Big( \int_{-\tilde{b}}^{y} \bar{\delta}_{y'} \tilde{\varrho}_f(y') dy'+ \int_{0_-}^{y+\bar{\delta}_y} \delta_{y'} \tilde{\varrho}_f(y') dy' \Big) \Big]-\tilde{\mu}_2 =0,
\label{eq:se_fless_edge_kmone_scaled_text}
\end{align}
where we have used the fact $\tilde{\ell} \to 0$ in the large $N$ limit.  In fact, we will later show that $\ell_N \sim \mathcal{O}(\sqrt{\log N})$.  In the above equation, $\bar{\delta}_y=\bar{\Delta}_{x=Ny}/N$ and $\delta_y=\Delta_{x=Ny}/N$ with $\Delta_x$ and $\bar{\Delta}_x$, representing the distances from point $x$, respectively,  on the left and right side over which one would find $f$ fraction of particles (see their definitions in Eq.~\eqref{Delta_x}). The chemical potential $\tilde{\mu}_2=\mu_2/N^3$ ensures that the density profile $\tilde{\varrho}_f(y)$ on the left edge is normalised to $f$.  One can write an equation similar to Eq.~\eqref{eq:se_fless_edge_kmone_scaled_text} for the right part as well. Owing to the symmetry between two parts, the calculation on the right part would be exactly same as in the left part. Therefore, here we only discuss the computations for the left part. Our next task is to insert  the ansatz from Eq.~\eqref{eq:density_k_mone_scaled_text} for the scaled density profile in Eq.~\eqref{eq:se_fless_edge_kmone_scaled_text} and evaluate different integrals in this equation. For that we first need to find the $y$ dependence of $\bar{\delta}_y$ and  $\delta_y$ explicitly. It turns out that the dependence of  these functions on $y$ changes as $f$ decreased below $1/3$. This happens because the number of particles $(1-2f)N$ in the central region changes from being smaller to being larger than the number of particles $fN$ at the edges as $f$ is decreased below $1/3$. Hence, one requires to consider the two cases $1/3 \leq f<1/2$ and $0<f<1/3$ separately.

We first discuss the  $1/3 \leq f<1/2$ case. The functions $\bar{\delta}_y$ and $\delta_y$ are determined in Eq.~\eqref{eq:delta_tilde_bar} and Eq.~\eqref{eq:delta_tilde} respectively and they are given by 
\begin{align}
 \bar{\delta}_y = 
 \begin{cases}
     -y ~&~~\text{for}~~-\tilde{b}\le y\le -\tilde{x}_f \\
     ~\tilde{x}_f ~ & ~~\text{for}~~-\tilde{x}_f \le y < 0_- \\
     ~~\tilde{b}  ~& ~~\text{for}~~~~~~~~~y=0_+,
 \end{cases}~~~,~~~
  \delta_y = 
 \begin{cases}
     y ~&~~\text{for}~~\tilde{x}_f \le y \le \tilde{b} \\
     \tilde{x}_f ~ & ~~\text{for}~~0_+< y \le \tilde{x}_f\\
     \tilde{b}  ~& ~~\text{for}~~~~~y=0_-,
 \end{cases}
 \label{eq:delta_y}
\end{align}
with 
\beq 
\tilde{x}_f = 
\frac{(3f-1)}{\tilde{\varrho}_{\rm edge}}~~\text{for}~\frac{1}{3} \leq f < \frac{1}{2}. \label{tx_f-txt}
\eeq
We use the above explicit expressions of the functions $\bar{\delta}_y$ and $\delta_y$ and the ansatz for $\tilde{\varrho}_f(y)$ from Eq.~\eqref{eq:density_k_mone_scaled_text} in the SPE \eqref{eq:se_fless_edge_kmone_scaled_text}. After performing the integrals in Eq.~\eqref{eq:se_fless_edge_kmone_scaled_text} we find the following equations  [details are provided in \ref{1docp-sd-1/3<f<1/2}] 
\begin{eqnarray}
\Big(&1- \tilde{\varrho}_{\text{edge}} \Big ) \frac{y^2}{2} \approx \tilde{\mu}_2, &~\text{for}~-\tilde{b}\leq y\leq -\tilde{x}_f, \notag \\
(&1- \tilde{\varrho}_{\text{edge}}) \frac{y^2}{2} +   \tilde{\varrho}_{\text{edge}} \Big( \frac{f}{\tilde{\varrho}_{\text{edge}}}-\tilde{b} \Big)  y - \frac{ \tilde{\varrho}_{\text{edge}}}{4}&\Big[\tilde{b}^2-\left ( \frac{3f-1}{\tilde{\varrho}_{\text{edge}}}\right )^2\Big] \label{eq:simplification_2-txt} \\
&~~~+\frac{1}{2} (1-2f) \Big[\tilde{b}-\frac{(1-2f)}{2\tilde{\varrho}_{\rm edge}} \Big] \approx \tilde{\mu}_2, &~ \text{for}~-\tilde{x}_f\leq y < 0_-.  \notag 
\end{eqnarray}

\noindent
Since the above equations are valid for arbitrary $y$ in the respective regions, the coefficients of different powers of $y$ should vanish individually in both the regions of $y$. Solving the resulting equations, we get
\begin{align}
 \tilde{\varrho}_{\rm edge} &=1,~~\tilde{b}=f,~~\text{and}~\tilde{\mu}_2=0.\label{consts-f>1/3}
\end{align}
 A similar calculations is performed for the $0<f<1/3$ case in \ref{1docp-sd-0<f<1/3} and once again one finds the same solutions for the constants  as in Eq.~\eqref{consts-f>1/3}. We would like to mention again that a similar calculation can be performed for the SPE on the right edge part and arrive at the same solutions as in Eq.~\eqref{consts-f>1/3}. Note that the  solutions in this equation completely specifies the expression of the scaled density profile in Eq.~\eqref{eq:density_k_mone_scaled_text} which is given explicitly in Eq.~\eqref{trho-k_-1-final-f<1/2} and verified numerically  in Fig.~\ref{smallf_kmone_figure}(a).

We now focus on the central regime which is described by the delta function part of the scaling distribution in Eq.~\eqref{eq:density_k_mone_scaled_text}. However, if one zooms over this region, then one does not find that $(1-2f)$ fraction of particles at the central region are  just piled up at the origin. They of course, get spread over a region $[-\ell_N,\ell_N]$ due to the presence of finite temperature, however as we will show below $\ell_N/N \to 0$ as $N \to \infty$. This spreading happens because of the entropy of these particles. As it turns out  the net interaction energy of the particles in this region is sub-dominant compared to the entropy.  Hence, neglecting the interaction part of the free energy, the SPE in the central region  can be written in terms of the un-scaled density profile  $\varrho_N(x)$  as 
\beq
N \frac{x^2}{2}+N\beta^{-1} \log[\varrho_N(x)]=\mu_1,
\label{eq:sp_free_text}
\eeq
which has the following solution 
\beq
\varrho_N(x)=\text{exp}\left(\frac{\mu_1}{N}-\beta \frac{x^2}{2}\right).
\label{eq:density_central_kmone_text}
\eeq
Note that the above Gaussian form of the density profile remains valid over the central region $[-\ell_N,\ell_N]$ and it should smoothly connect to the density profile outside the central region on both sides. Furthermore, $\mu_1$ should be such that the fraction of the number of particles inside the central region is $(1-2f)$. These conditions provide the following two equations 
\begin{align}
\text{exp}\left(\frac{\mu_1}{N}-\beta \frac{\ell^2_N}{2}\right)&=\frac{1}{N}, 
\label{eq:continuity_text} \\
\text{with}~~\int_{-\ell_N}^{\ell_N} \varrho_N(x) dx&=(1-2f). 
\label{eq:normalization_central_text}
\end{align}
Solving these two equations one finds that 
\begin{align}
\ell_N\approx \sqrt{\frac{2}{\beta}\log N},~~\text{and}~ \exp(\mu_1/N) \approx (1-2f)\sqrt{\frac{\beta}{2 \pi}}, \label{l-sim-sq-logN}
\end{align}
which completely specify the density profile in Eq.~\eqref{eq:density_central_kmone_text} in the central region $[-\ell_N,\ell_N]$. This Gaussian form for the density profile is verified in Fig.~\ref{smallf_kmone_figure}(b).

In summary, for fixed $f \in [\frac{1}{2},1]$ and large $N$, the equilibrium density profile is piece-wise uniform 
	\beq
	\varrho_N(x) \approx \begin{cases}
	        & \frac{1}{N} \text{~~~for~~} x \in\big[-fN,-(2f-1)N \big), \\
		& \frac{1}{2N} \text{~~for~~} x \in \big[-(2f-1)N,(2f-1)N \big] \\
		& \frac{1}{N} \text{~~~for~~} x \in \big ((2f-1)N, fN \big],
		\label{density_kmone_fgreater}
	\end{cases}
	\eeq
On the other hand for $0<f<1/2$, the equilibrium density profile has the following form
	\beq
	\varrho_N(x) \approx 
	\begin{cases}
	~~~~~~~~~~ \frac{1}{N}& \text{~for~~} x \in \big [-fN ,-\ell_N\big).\\
		(1-2f) \sqrt{\frac{\beta}{2\pi}}e^{-\beta x^2/2} &\text{~for}~~x \in [-\ell_N,\ell_N],~\text{with}~\ell_N \sim \mathcal{O}(\sqrt{\log N}) \\
		~~~~~~~~~~ \frac{1}{N}& \text{~for~~} x \in \big (\ell_N, fN \big],
	\end{cases}
	\label{density_kmone_fless}
	\eeq
	for large $N$. In terms of the scaled variable $y$, the density profiles in the above equations posses the scaling form in Eq.~\eqref{den-sc-form-fr}  with $\alpha=1$ and the scaling functions are given in Eq.~\eqref{trho-k_-1-final-f>1/2} and Eq.~\eqref{trho-k_-1-final-f<1/2}. These  scaling functions can indeed describe the crossover demonstrated in Fig.~\ref{density_summary}(a). Note, for $f=1$ these results reproduce the density profile of the ATAC case given in Eq.~\eqref{sc-den-f-1} which represents a flat profile existing over a single support $[-N,N]$.  As $f$ is reduced from $1$, the particles at the edges do not feel the repulsion from all the particles, however the particles at the central part feel the interaction from all other particles.  Consequently,  the total support of the profile shrinks to $[-fN,fN]$ and the density profile becomes piecewise uniform over three parts. Since the particles in the central part feel more repulsion than the edge parts, more particles get pushed away from the centre of the trap and giving  rise to higher density value in the edge parts than the central part.  As $f$ is further reduced, the two edge parts approach each other, causing the central part to shrink and cease to exist at $f=1/2$. When $f$ is reduced below $1/2$, one does not find a single particle in the  system which interacts with  all other particles. On the other hand, because of the reduced $f$, the amount of overall repulsion that the edge particles can manage  to climb against the harmonic potential also got reduced. Hence they get more squeezed leading to an even smaller spread ($2fN$) for the full profile. In this situation to minimize energy, the particles at the central part find themselves easier to sit on top of each other at the minimum of the harmonic trap leading to piling up at the centre. At this point the entropy starts dominating because the contribution from interaction part of the energy to the free energy becomes negligible. As a result the particles behave like a non-interacting fluid and gets distributed according to the Boltzmann distribution which is a Gaussian in harmonic trap. With decreasing $f$, more and more particles join the fluid in the central part. In the $f \to 0$ limit the edge parts get completely melted into the fluid part  and one finds the density profile to be Gaussian as was obtained in Ref:~\cite{Avanish_PRE_2020}.


\section{Density crossover in log-gas ($V(x) = -\log|x|$).}
\label{crossover_genkgt0}
The energy function of log-gas is given in Eq.~\eqref{E-ocp-lg}. In the ATAC ($f=1$) case , as mentioned earlier, the equilibrium density profile has the Wigner semi-circular scaling form given in Eq.~\eqref{sc-den-f-1} \cite{sanaa2019}.  On the other hand, in the SR case ($f=0$), the equilibrium density profile is  Gaussian and the width of the profile depends on the parameter $d$ (see Eq.~\eqref{eq:denk<0}) \cite{Avanish_PRE_2020}. As demonstrated in Fig.~\ref{density_summary}(b), the density profile changes from Wigner semi circle to Gaussian as $f$ is decreased. To understand this crossover in the equilibrium density profile one requires to solve the relevant SPEs for arbitrary $f \in [0,1]$. While exact solutions for the scaling density function $\tilde{\varrho}_f(y)$ have been obtained for $f=1$ and $f=0$, finding the solution for general $f$ seems difficult. However, it turns out that for $f$ close to $1$ and $0$, one can find approximate solutions to the SPEs, which we discuss below.

\subsection{Solution for $f$ close to $1$:}
\noindent For $1/2 \le f \le 1$ the SPEs in the central, left and right edge parts are derived in Eqs.~(\ref{SPE_genk_f>1/2_center} - \ref{SPE_genk_f>1/2_right}). Here, we write them explicitly with log-gas interaction $ V(x)=-\log(|x|)$. In the central part $-\ell_N \le  x \le \ell_N$ one has 
\begin{align}
N \frac{x^2}{2}-&  N^2~ \int_{-b_N}^{b_N}  \log(|x'-x|)~\varrho_N(x')~ dx' \notag \\
& +\frac{N^2}{2} \Big[ \int_{-b_N}^{-\ell_N}  \log(\bar{\Delta}_{x'})~\varrho_N(x')~ dx' + \int_{\ell_N}^{b_N} \log(\Delta_{x'})~\varrho_N(x') ~ dx' \Big] =\mu_1,
\label{SPE_genk_flarge_center}
\end{align}
and in the left part ($-b_N \le x<-\ell_N$) one has
\begin{align}
N\frac{x^2}{2} -N^2~  \int_{-b_N}^{b_N}  & \log(|x'-x|)~\varrho_N(x')~ dx'+\frac{ N^2}{2} \left[2 \int_{x+\bar{\Delta}_x}^{b_N}  \log(|x'-x|)\varrho_N(x')dx' \right. \notag \\
& \left. +  \int_{-b_N}^{x} \log(\bar{\Delta}_{x'}) \varrho_N(x')dx' +\int_{\ell_N}^{x+\bar{\Delta}_x} \log(\Delta_{x'})\varrho_N(x')dx' \right]=\mu_2.
\label{SPE_genk_flarge_edge}
\end{align}
One can write a similar equation on the right part, which has the same structure due to the inversion symmetry (about the origin) of the problem.
The chemical potentials $\mu_1$ and $\mu_2$ ensure the two normalisation constraints given in  Eq.~\eqref{constraint_fgt1}. Recall that $\Delta_x$ ($\bar{\Delta}_x$), defined in Eq.~\eqref{Delta_x}, represents the distance from $x$ over which one would find $f$ fraction of particles on the left (right) side.

\begin{figure}[t]
	\centering		
	\includegraphics[width=1.0\textwidth]{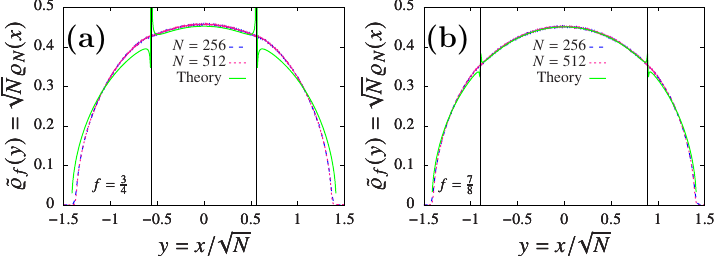}	
	\caption{These plots show system size collapse of the numerical density profiles (dashed lines) for different $N$ in the log-gas model for (a) $f=\frac{3}{4}$  and (b) $f=\frac{7}{8}$. The good collapse verifies the scaling $\varrho_N(x)= \frac{1}{\sqrt{N}} \tilde{\varrho}_f(x/\sqrt{N})$ in Eq.~\eqref{den-sc-form-fr}. The solid green line plots the scaling function $\tilde{\varrho}_f(y)$ as given in Eq.~\eqref{density_index_satya}. We observe good agreement between simulation and theory in the central regime (represented by solid black vertical lines at $y=\pm \tilde{\ell}$). Outside this regime, the agreement between theory and simulation is better for $f$ close to one and declines as $f$ is reduced  from one.} 
	\label{largef_kzero_figure}
\end{figure}

For $f$ close to $1$, most of the particles stay in the central region and consequently, the edge regions being lightly populated become much narrower than the central region. Hence, in this limit, the contribution from the third terms (inside the square brackets) on the L.H.S. of Eq.~\eqref{SPE_genk_flarge_center} and Eq.~\eqref{SPE_genk_flarge_edge} are much smaller than the first and second terms. We neglect these terms and solve the resulting approximate SPEs which look like 
\begin{align}
\begin{split}
N \frac{x^2}{2}&-N^2~ \int_{-b_N}^{b_N}  \log(|x'-x|)~\varrho_N(x')~ dx'\approx \mu_1,~~\text{for}~~|x| \le  \ell_N \\
N\frac{x^2}{2} &-N^2~  \int_{-b_N}^{b_N}  \log(|x'-x|)~\varrho_N(x')~ dx' 
          \approx \mu_2,~~~\text{for}~\ell_N<|x| \le b_N,
          \end{split}
\label{0-order-0<k<1}
\end{align}
with the constraints $\int_{-\ell_N}^{\ell_N} \varrho_N(x)~dx=2f-1$ and $\int_{\ell_N < |x| \le b_N} \varrho_N(x)~dx=1-f$.
Note that the saddle point problem in Eq.~\eqref{0-order-0<k<1} is equivalent to the problem of computing the equilibrium density profile in an ATAC log-gas system, where $(2f-1)N$ number of particles are restricted to be inside $[-\ell_N,\ell_N]$ and the rest of the particles stay outside. For log-gas, this problem was solved in Ref.~\cite{Phase_PRL_marino} subject to the constraint of having a fixed  number of particles inside $[-\ell_N,\ell_N]$. The density profile is  given by $\varrho_N(x)=(1/\sqrt{N})\tilde{\varrho}_f(x/\sqrt{N})$ with $\tilde{\varrho}_f(y)$ given explicitly in Eq.~\eqref{density_index_satya} with $\tilde{\ell}=\ell_N/\sqrt{N}$ and $\tilde{b}=b_N/\sqrt{N}$.
We adopt this solution in our case however,  $\ell_N$ is not known beforehand.  We calculate its numerical value, $\ell_{\rm nu}$ using $\int_{-\ell_{\rm nu}}^{\ell_{\rm nu}}\varrho_{\rm nu}(x) dx=(2f-1)$, where $\varrho_{\rm nu}(x)$ being the numerically obtained density profile.
The analytical form in Eq.~\eqref{density_index_satya} is compared with the numerically obtained densities in Fig.~\ref{largef_kzero_figure}, where we notice good agreement between them. Closer the $f$ is to one, better  the agreement is.

\subsection{Solution for $f$ close to $0$:}
\label{sec:0<k<1-f<1/2}
\noindent 
The SPEs for $0<f<1/2$ are different from the previous case and are derived in Eqs.~(\ref{SPE_genk_f<1/2_left} - \ref{SPE_genk_f<1/2_center}) for the three regimes, where one needs to insert $V(x)=-\log(|x|)$. The SPE in the central part $-\ell_N \leq x \leq \ell_N$ is  written explicitly as  
\begin{align}
\frac{Nx^2}{2}-&\frac{N^2}{2} \Big[2 \int_{x-\Delta_x}^{x+\bar{\Delta}_x} ~ \log(|x'-x|) -\int_{x-\Delta_x}^{x} \log(\bar{\Delta}_{x'})-\int_{x}^{x+\bar{\Delta}_x} \log(\Delta_{x'}) \Big] ~\varrho_N(x') dx'\notag \\
& -\frac{N^2}{2}
\int_{-b_N}^{-\ell_N}    \Big[ \log(|x'-x|) -  \log(\bar{\Delta}_{x'}) \Big] \Theta(x'+\bar{\Delta}_{x'}-x)~ \varrho_N(x') ~dx' 
\label{saddle_kgreater_fless} \\		
& -\frac{N^2}{2} \int_{\ell_N}^{b_N}  \Big[ \log(|x'-x|) -  \log(\Delta_{x'}) \Big] \Theta(x-x'+\Delta_{x'}) ~ \varrho_N(x')~dx'=\mu_1, \notag 
\end{align}
where $\Theta(x)$ is the Heaviside step function.

We, here, solve this equation approximately for small $f$.  Similar to the previous case ($f$ close to $1$), in this case also the central region $[-\ell_N,\ell_N]$ is much wider than the two edge regimes  [$-b_N,-\ell_N$) and  ($\ell_N,b_N$]. 
Recall the central region contains $(1-2f)$ fraction of particles and the two edge regions contain $f$ fractions of particles each. As $f$ approaches to zero, the central region widens and the edge regions shrink. 
Starting from any point $x' \in [-b_N,-\ell_N)$ one must traverse to the image point $\bar{X}(x')=x'+\bar{\Delta}_{x'}$ on the 
right of  $x'$, to find $f$ fraction of particles. For $f$ close to $0$, one does not require to move far inside the bulk.  
More precisely, the distance $\bar{\Delta}_{-\ell_N}=\big{(}\bar{X}(-\ell_N)+\ell_N\big{)}$ is very small for small $f$ and $\bar{X}(x')<\bar{X}(-\ell_N)$  for any $x'  \in [-b_N,-\ell_N)$.  Additionally, the presence of $\Theta(x'+\bar{\Delta}_{x'}-x)$ on the second line of Eq.~\eqref{saddle_kgreater_fless} makes the contribution from this integral non-zero only over a tiny region $-\ell_N \leq x \leq \bar{X}(-\ell_N)$. 
For $x$ in this small region, we neglect the contribution from the integral on the second line of Eq.~\eqref{saddle_kgreater_fless}.  
Following similar arguments, the contribution from the integral on the third line of Eq.~\eqref{saddle_kgreater_fless} is also ignored in the $f \to 0$ limit. As a justification, we later show that the resulting approximate solution for the density profile in the central part indeed matches quite well with the numerical results.

\begin{figure}[t]
\centering		
\includegraphics[width=1.0\textwidth]{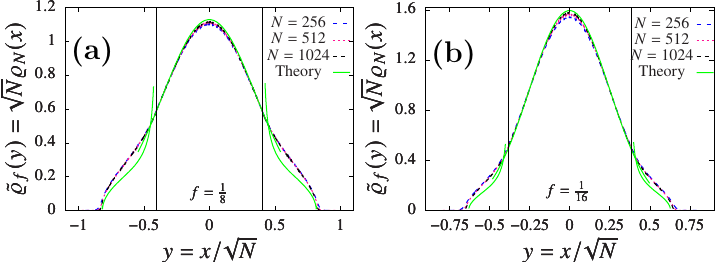}		
\caption{Comparison of numerically obtained (scaled) density profiles for different system sizes (dashed lines) with the analytical predictions (green solid lines) for log-gas  (a) with $f=1/8$ and (b) with $f=1/16$. Once again collapse of numerical data for different $N$ verifies the scaling in Eq.~\eqref{den-sc-form-fr} with $\alpha=1/2$. The  theoretical prediction for the form of the scaled density profile $\tilde{\varrho}_f(y)$ inside the central part $(-\tilde{\ell},\tilde{\ell})$  is Gaussian as in  Eq.~\eqref{smallf_kzero_theory} [for log-gas]. The central region $(-\tilde{\ell},\tilde{\ell})$ (represented by two vertical solid black lines) can be found from the relation $\int_{-\tilde{\ell}}^{\tilde{\ell}} \tilde{\varrho}_f(y) dy =1-2f$. Outside the central region, the equilibrium density profile $\tilde{\varrho}_f(y)$ is different from the central part and its approximate analytical formula is provided in  Eq.~\eqref{density_edge_log}. Although agreement between the analytical formula [Eq.~\eqref{density_edge_log}] for the density and the numerical density profile reduces as $f$ is increased from  zero, however, it captures the total support of the density profile quite accurately.}
\label{smallf_kzero_figure}
\end{figure}

Neglecting the contributions from these integrals on the second and third lines of Eq.~\eqref{saddle_kgreater_fless}, the SPE in now  approximately reads

\begin{align}
\frac{Nx^2}{2}-&N^2 \int_{x-\Delta_x}^{x+\bar{\Delta}_x} ~ \log(|x'-x|)\varrho_N(x')~dx' \notag \\
&+\frac{N^2}{2} \int_{x-\Delta_x}^{x} \log(\bar{\Delta}_{x'})\varrho_N(x')~dx'+\frac{N^2}{2}\int_{x}^{x+\bar{\Delta}_x} \log(\Delta_{x'}) \varrho_N(x')~dx' \approx \mu_1.
\label{eq:SPE_fless_simplifies}
\end{align}
For small $f$,  $\Delta_x$ and $\bar{\Delta}_x$ can be approximated by $\frac{f}{\varrho_N(x)}$  which can be obtained from  Eq.~\eqref{Delta_x}. After a few simple algebraic steps, one can calculate the contribution from interaction part of the energy in the SPE. Apart from some numerical additive constants, which are independent of $x$, the contribution of the interaction term comes out to be of order $ f N^2 \log[\varrho_N(x)]$.  The SPE then becomes
\beq
\frac{Nx^2}{2}+f N^2 \log(\varrho_N(x)) \approx \mu_1.
\label{SPE_effective}
\eeq
Solving this equation one finds 
\begin{align}
\begin{split}
\varrho_N(x) &\approx A_N \exp\left(-\frac{x^2}{2Nf}\right) 
~~\text{for}~~-\ell_N \le x \le \ell_N, \\ 
\text{where}~&~A_N =\exp\left( \frac{\mu_1}{N^2 f} \right).
\end{split}
\label{smallf_kzero_theory-2}
\end{align}
This density profile should satisfy the normalisation  $\int_{-\ell_N}^{\ell_N}\varrho_N(x) dx=1-2f$ which provides a relation between $\mu_1$ and $\ell_N$ for given $f$. It is easy to realise that in the $f \to 0$ limit, the support length $\ell_N \to \infty$. Using this fact, we find that 
$A_N \approx 1/\sqrt{2\pi Nf}$ {\it i.e.,} $\mu_1\approx -(N^2 f/2)\log(2\pi N f)$ in the leading order in $f$. Making use of this approximate expression of $A_N$ back in the normalisation condition yields $\text{erf}(\ell_N/\sqrt{2Nf}) \approx 1-2f$,  solving which one can find 
$\tilde{\ell}=\ell_N/\sqrt{N}$ in the leading order in $f$. Note that the density profile in the central regime $-\ell_N \leq x \leq \ell_N$ given in Eq.~\eqref{smallf_kzero_theory-2} has the scaling form $\varrho_N(x)=(1/\sqrt{N})\tilde{\varrho}_f(x/\sqrt{N})$ with $\tilde{\varrho}_f(y)$ given explicitly in Eq.~\eqref{smallf_kzero_theory}. A comparison of this result with numerical simulation is provided in Fig.~\ref{smallf_kzero_figure} for $f=1/8$ and $f=1/16$. The excellent agreement of theory with the numerical profiles in the central part provides justification of the approximations made to go from Eq.~\eqref{saddle_kgreater_fless} to Eq.~\eqref{eq:SPE_fless_simplifies}. 

The density profile in Eq.~\eqref{smallf_kzero_theory-2} does not explain the behaviour of the density of the remaining $2fN$ particles in the left and right edge parts. These regimes therefore need to be considered separately. We first focus on the left edge part.
The SPE in the left edge part ($-b_N \le x<-\ell_N$) can be written from Eq.~\eqref{SPE_genk_f<1/2_left} with $V(x)=-\log(|x|)$ as 
\begin{align}	
\frac{Nx^2}{2}-N^2 \int_{-b_N}^{-\ell_N}  &\log(|x'-x|) ~\varrho_N(x')~dx'  ~-\frac{N^2}{2}\left[2 \int_{-\ell_N}^{x+\bar{\Delta}_x}  \log(|x'-x|)\varrho_N(x')dx' \right. \label{eq:se_fless_edge} \\
 &\left. - \int_{-b_N}^{x} \log(\bar{\Delta}_{x'})\varrho_N(x')dx'- \int_{-\ell_N}^{x+\bar{\Delta}_x}  \log(\Delta_{x'})\varrho_N(x')dx'\right]  =\mu_2. 
 \notag
\end{align}
It turns out that this SPE is in general difficult to solve for arbitrary $f \in (0,1/2)$. We however notice that for small $f$, the contribution from the last three terms on the L.H.S. of Eq.~\eqref{eq:se_fless_edge} ({\it i.e.} the terms inside the square bracket) should be much smaller than the second term (in the first line). 

As mentioned previously,  the size $(b_N-\ell_N)$ of  an edge part is very small for small $f$ and consequently, for any $x \in [-b_N,-\ell_N)$ the image point $(x+\bar{\Delta}_x) \approx -\ell_N$. Hence, the integration ranges of the three terms inside the square bracket on the L.H.S. of Eq.~\eqref{eq:se_fless_edge} are very small and one can neglect them compared to the first two terms. This approximation is exact for $x=-b_N$ and it becomes less valid as $x$ is taken away from $-b_N$. Owing to the symmetry of the problem, one can make the same set of approximations for the SPE on the right edge part  $\ell_N<x \le b_N$. After making the above mentioned approximations, the SPEs on both edge parts now read
\begin{align}
\begin{split}
\frac{Nx^2}{2}&-N^2 \int_{-b_N}^{-\ell_N}  \log(|x'-x|)~\varrho_N(x')~dx' \approx \mu_2, ~~\text{for}~~-b_N \le x<-\ell_N  \\
  \frac{Nx^2}{2}&-N^2 \int_{\ell_N}^{b_N}  \log(|x'-x|)~\varrho_N(x')~dx' \approx \mu_2, ~~~~\text{for}~~~~~\ell_N<x \le b_N. 
\end{split}
\label{free_equation}
\end{align}

\noindent 
The problem in Eq.~\ref{free_equation} can be interpreted as the problem of finding the density profile of the $2fN$ particles interacting all-to-all and conditioned to be outside the central range $[-\ell_N,\ell_N]$. This problem was studied in Ref.\cite{Phase_PRL_marino}. For this case, the scaled density profile $\tilde{\varrho}_f(y)$ takes the same form as in Eq.\eqref{density_index_satya} with $\tilde{a}=0$, since the central region $[-\ell_N,\ell_N]$ is empty, and thus is given by Eq.\eqref{density_edge_log}. The analytical form in Eq.~\eqref{density_edge_log} for the equilibrium density at the edges is compared with the density obtained numerically in Fig.~\ref{smallf_kzero_figure}. The support of the density $\tilde{b}$ can be computed from normalisation and one finds  $\tilde{b}=\sqrt{\tilde{\ell}^2+4f}$ which quite accurately describes the support of the simulation density profile in Fig.~\ref{smallf_kzero_figure} even though the analytical form in Eq.~\eqref{density_edge_log} is only approximate. This happens because Eq.~\eqref{free_equation} is exact only when $|x|=b_N$ and the approximations for getting Eq.~\eqref{free_equation} starting from Eq.~\eqref{eq:se_fless_edge} does not hold for $x$ away from the edges.

In summary, we found approximate forms of the equilibrium density profiles of the FR log-gas model for $f$ close to $1$ given by Eq.~\eqref{density_index_satya} and for small $f$ given by Eqs.~\eqref{smallf_kzero_theory} and \eqref{density_edge_log}. These theoretical results are compared with simulation data in Fig.~\ref{largef_kzero_figure} and Fig.~\ref{smallf_kzero_figure} respectively. We observe that the density profiles for large and small $f$ are quite different. From Fig.~\ref{largef_kzero_figure}, with $f$ decreasing from one, the edge parts grow inside the central part as in the $1$dOCP case discussed in the previous section. At $f=1/2$  the central part vanishes. On the other hand, for small $f$, we observe from Fig.\ref{smallf_kzero_figure}(a) (for $f=1/8$) and Fig.\ref{smallf_kzero_figure}(b) (for $f=1/16$) that with decreasing $f$, the density profile in the central part grows, causing the edge parts to vanish at $f=0$. At this value of $f$,  the scaled density profile $\tilde{\varrho}_f(y)$ is described by Eq.~\eqref{smallf_kzero_theory} and reproduces the result obtained previously in Ref:~\cite{Avanish_PRE_2020}. 

\begin{table}[h]
\centering
\begin{tabular}{| c | c | c | c | c | c |}
	\hline 
	\diagbox[width=\dimexpr \textwidth/11+8\tabcolsep\relax, height=1.cm]{\bf{Features}}{\bf{Model}}
	& \multicolumn{2}{c} {\textbf{{$\bm{1}$dOCP ($\bm{k=-1}$)}}} & \multicolumn{2}{|c|} {\textbf{{log-gas ($\bm{k \to 0}$)}}} & $\substack{\text{\large{other}} \\ \Large{k>-2}}$ \\
	\hline 
 & \multicolumn{2}{c} {Numerically } \vline  & \multicolumn{2}{c} {Numerically} \vline& { Numerically}\\ 
	\textbf{{Crossover}}	& \multicolumn{2}{c} {{demonstrated}} \vline  & \multicolumn{2}{c} { demonstrated} \vline&{demonstrated}\\ 
	&  \multicolumn{2}{c}{\centering {in Fig.~\ref{density_summary}(a)}} \vline & \multicolumn{2}{c}{\centering {in Fig.~\ref{density_summary}(b)}} \vline & {in Fig.~\ref{fig:density_crossover}}   \\
	\cline{1-6}
	& \multirow{1}{*}{\centering \textbf{Entropy}} &   {\centering {Subdominant$^*$}} & \multicolumn{2}{c} {Subdominant} \vline  &  \\ 
	\cline{2-5}
	\textbf{{Free}}  & {\centering \textbf{Bulk}} &   \multirow{1}{*}{\centering {Dominant}} & \multicolumn{2}{c} {{Dominant}} \vline  &  \\ 
	\textbf{{energy}}  & {\centering \textbf{energy}} &   \multirow{1}{*}{\centering {contribution}} & \multicolumn{2}{c} {{ contribution}} \vline  & \\ 
	\cline{2-5}
	\textbf{{functional}} & {\centering \textbf{Self}} &   \multirow{1}{*}{\centering {No}} & \multicolumn{2}{c} {{Neglected for}} \vline  &  \\ 
	& {\centering \textbf{energy}} &   \multirow{1}{*}{\centering {contribution}} & \multicolumn{2}{c} {{fixed $f>0$}} \vline  & Only  \\ 
	\cline{1-5}
	\textbf{{Saddle point}}  & \multicolumn{2}{c}{ \centering {Derived in }} \vline & \multicolumn{2}{c} {\centering {Derived in }} \vline & Numerical  \\
	\textbf{{equation}}  & \multicolumn{2}{c}{ \centering {\ref{sadle-eq-derv}}} \vline & \multicolumn{2}{c} {\centering {\ref{sadle-eq-derv}}} \vline & observation \\
	\cline{1-5}
	& \multicolumn{2}{c}{\centering {Exact analytical }} \vline &  \multicolumn{2}{c}{\centering {Approximate }} \vline & \\
	& \multicolumn{2}{c}{\centering {expression for $\tilde{\varrho}_f(y)$}} \vline &  \multicolumn{2}{c}{\centering {expression for $\tilde{\varrho}_f(y)$}} \vline & \\
	\cline{2-5} 
	& & {Given in} &  & {Given in } & \\
	&{$\bm{f \in }$}  & {Eq.~\ref{trho-k_-1-final-f>1/2}} & $\bm{f}$ & {Eq.~\ref{density_index_satya}} & \\
	\cline{3-3} \cline{5-5}
	& {$\bm{\big{[}\frac{1}{2},1\big{]}}$}& {Verified } &{\bf{close }}  & {Compared} & \\
	& & {numerically } &  \bf {to} $\bm{1}$  & {with numerics} & \\
	\textbf{{Scaling }}  & & {in Fig.~\ref{largef_kmone_figure}} & & {in Fig.~\ref{largef_kzero_figure}} & \\
	\cline{2-5}
	\textbf{{function }}  & & {Given in } & & {Given in }  & \\
	{$\bm{\tilde{\varrho}_f(y)}$} & {$\bm{f \in }$}  & Eq.~\ref{trho-k_-1-final-f<1/2}  & $\bm{f}$ & {Eq.~\ref{smallf_kzero_theory}} \& \ref{density_edge_log} & \\
	\cline{3-3} \cline{5-5}
	& {$\bm{\big{(}0,\frac{1}{2}\big{)}}$}  & {Verified} &  {\bf close}  & {Compared} & \\
	& & {numerically} &  {\bf{to} $\bm{0}$} & {with numerics} & \\
	& & {in Fig.~\ref{smallf_kmone_figure}} &  & {in Fig.~\ref{smallf_kzero_figure}} & \\
	\cline{1-6}
\end{tabular}
\caption{A table summarising the status of our results. The $^*$ in the second row indicates the fact that  for $0<f<1/2$ the entropy can contribute in the  central part of the density profile in the $1$dOCP model. However the central region $[-\ell_N,\ell_N]$ shrinks to a point in terms of scaled variable $y=x/N$ in the $N \to \infty$ limit since $\ell_N \sim \sqrt{\log N}$ [see Eq.~\eqref{l-sim-sq-logN}].}
\label{table2}
\end{table}

\vspace{1cm}
	\section{Conclusions and outlook}
	\label{conclusions}
We  studied equilibrium density profiles of two harmonically confined classical many particle model systems -- $1$dOCP  and log-gas, in which the particles are interacting via potentials $V(x) =-|x|$ and $V(x)=-\log(|x|)$, respectively. Additionally, the interaction is such that each particle, in both systems, can interact only upto $d$ particles to its left and right if available.  We numerically demonstrated that as the parameter $f=d/N$ is reduced from $1$ to $0$, the density profile in both models undergo crossover from a finitely supported density profile to an infinitely extended Gaussian density profile. Furthermore, for both the models we found that for all $f \in (0,1]$ the density profile possess same scaling with respect to the systems size $N$ as in their respective ATAC cases. To understand the above mentioned  crossover of the density profile, we have derived the corresponding saddle point equations for both models. We found that the SPEs are different for $1/2\leq f\le1$ and $0 < f < 1/2$. For the $1$dOCP model, we solved the SPEs in all regimes of $f$ and found exact analytical expressions for  the density scaling functions which are verified with numerical computations. For the log-gas, we provided approximate solutions of the SPEs for large (close to $1$) and small (close to $0$) $f$.  We have numerically found  similar crossover in the density profiles to exist for other Riesz gas models corresponding to other values of $k$ in Eq.~\eqref{gen_model}. A brief discussion along with numerical demonstration of such crossover for other values of $k$ is provided in \ref{app:den-cross-other-k}. 
We provide a table of summary in Tab.~\ref{table2} which contains essential features of our results along with the equation numbers and the section numbers where they appear or they are derived.

Our work can be extended in several direction in future. One immediate direction would be to find complete solution of the SPEs for arbitrary $f$ in the log-gas model as well as for other values of $k$ in Eq.~\eqref{gen_model}. 
It would be interesting to look at how the distribution of position of the particle at the edge gets affected as $f$ is reduced from one. In the all-to-all coupled case it was found that the spacing distribution exhibits interesting statistical fluctuations for different $k$~\cite{saikat_PRL_2022}.  It would also be interesting to investigate how the spacing statistics get affected as $f$ is reduced. Another interesting outlook would be to look at the crossover in densities due to temperature variation. Particularly for the ATAC log-gas ($f=1$), Gauss-Wigner crossover is observed in Ref.~\cite{Invariant_PRL_Allez} but how would such temperature tuned crossover happen for $f<1$? This is an interesting question to study.  Extensions of our study to higher dimensional Riesz gas is another interesting and challenging problem~\cite{Serfaty_arxiv_2017,Lewin_JSP_2022}. Exploring crossover in densities within quantum systems exhibiting power-law interactions could also be an intriguing avenue to investigate~\cite{Sassetti_PRB_2016,Sassetti_crystal_2021}.

\section{Acknowledgements}
We thank Manas Kulkarni for helpful discussions. 
A.K. acknowledges the support of the core research grant CRG/2021/002455 and the MATRICS grant MTR/2021/000350 from the SERB, DST, Government of India. SS and AK acknowledge support of the Department of Atomic Energy, Government of
India, under Project No. RTI4001.

 \appendix

\section{Numerical details}
\label{numerical_details}
In this section we provide details of our numerical computation. To determine the density profiles we compute the histograms of particle positions obtained through Monte-Carlo simulations at inverse temperature $\beta=1$. For each case the system is allowed to thermalise over  $10^6$ Monte-Carlo  cycles initially. By one Monte-Carlo  cycle, we mean $\mathcal{O}(N)$ Metropolis steps. We  perform average over $10^7$ samples to determine the density profiles.

\section{Field theoretic description of FR Riesz gas and the derivation of the saddle point equations}
\label{sadle-eq-derv}
In this section, we provide the derivation of the saddle point equations both for $1$dOCP and log-gas described by the energy functions given in Eq.~\eqref{E-ocp-lg}. As mentioned in the beginning of Sec.~\ref{quantity_summary}, for both the models with $0<f<1$ the support of the equilibrium density profile should consist of three regimes over which they behave differently. Such regimes naturally appear because the interaction felt by a particle in these three regimes are different since the number of other particles available to which it can interact, is different. Recall, we denote these regimes as central part, left edge part and the right edge part which are supported over the following regions $[-\ell_N,\ell_N]$, $[-b_N,-\ell_N)$ and $(\ell_N,b_N]$, respectively, with $0<\ell_N<b_N$. On an average, the central part contains $|1-2f|$ fraction of particles and the two edge parts contain rest of the fraction of particles equally divided. Consequently, the free energy functional defined in Eq.~\eqref{Sigma_mu} now reads 
\begin{align}
\bar{\Sigma}&[\rho_N(x)]=  \Sigma[\rho_N(x)]-\mu_1\left( \int_{-\ell_N}^{\ell_N} dx \rho_N(x) -|1-2f|\right) \\
&-\mu_2\left( \int_{-b_N}^{-\ell_N} dx \rho_N(x) -\frac{(1-|1-2f|)}{2}\right)-\mu_3\left( \int_{\ell_N}^{b_N} dx \rho_N(x) -\frac{(1-|1-2f|)}{2}\right),
\notag 
\end{align}
where $\mu_1,\mu_2$ and $\mu_3$ are the chemical potentials that ensure the following normalisations
\begin{align}
\begin{split}
\int_{-\ell_N}^{\ell_N} \rho_N(x)~dx&=|1-2f| ~~\text{and}  \\
\int_{-b_N}^{-\ell_N} \rho_N(x)~dx&=\int_{\ell_N}^{b_N} \rho_N(x)~dx=\frac{(1-|1-2f|)}{2},
\end{split}
\label{constraint_fgt1}
\end{align}
 in the central, the left edge and the right edge parts respectively. Due to the inversion symmetry of the microscopic energy function $\mathscr{E}(\{x_i\})$ in Eq.~\eqref{E-ocp-lg}, we expect the equilibrium density profile to be symmetric about the origin (the centre of the trap) which implies $\mu_2=\mu_3$ in equilibrium.
 
The free energy functional defined in Eq.~\eqref{eq:Sigma} is a combination of energy and entropy functionals $\Sigma[\rho_N(x)]=\mathcal{E}[\rho_N(x)] - \beta^{-1} \mathcal{S}[\rho_N(x)]$.
The entropy $\mathcal{S}[\rho_N(x)]$ can be computed by  `counting' the number of microscopic configurations compatible with a given density profile $\rho_N(x)$ which essentially provides $\mathcal{S}[\rho_N(x)]=-N\int dx \rho_N(x)\log \rho_N(x)$\redw{\cite{Dean_PRE_2008,dyson_JKP_1962}}. To find an expression for the energy functional $\mathcal{E}[\rho_N(x)]$ corresponding to the energy function in Eq.~\eqref{E-ocp-lg}, we first notice that the functional $\mathcal{E}[\rho_N(x)]$ has the form
\begin{align}
\mathcal{E}[\rho_N(x)] = \frac{N}{2}\int_{-b_N}^{b_N} x^2 \rho_N(x) dx + \mathcal{E}_I[\rho_N(x)], \label{free_energy_a}
\end{align}
where the first term comes from the confining potential in Eq.~\eqref{E-ocp-lg} and the second term $ \mathcal{E}_I[\rho_N(x)]$ represents contribution from the interaction potential. To find the equilibrium density profile for large $N$, one needs to solve the following saddle point equation 
\begin{align}
\frac{\delta \bar{\Sigma}[\rho_N(x)]}{\delta \rho_N(x)}\big{|}_{\rho_N(x)=\varrho_N(x)}=0,
\label{eq:spe2}
\end{align}
subject to the normalisation conditions in Eq.~\eqref{constraint_fgt1} which arise by equating the derivatives of $\bar{\Sigma}[\rho_N(x)]$ with respect ot $\mu_1$ and $\mu_2$ to zero, respectively.

As mentioned in Sec.~\ref{quantity_summary}, we numerically observe [from Figs.~(\ref{largef_kmone_figure}, \ref{smallf_kmone_figure}) for $1$dOCP and Figs.~(\ref{largef_kzero_figure}, \ref{smallf_kzero_figure}) for log-gas] that  the density profile possesses scaling form (see Eq.~\eqref{den-sc-form-fr})with a $N$-dependent length scale $L_N \sim N^{\alpha}$ where $\alpha$ is given in Eq.~\eqref{density_all_to_all}. Using this scaling form in the expression of the free energy, it is easy to notice that the energy $\mathcal{E}[\rho_N(x)]\simeq N\int_{-b_N}^{b_N}  (x^2/2)~\rho_N(x)~dx + \mathcal{E}_I[\rho_N(x)]$ scales as $N^{2\alpha+1}$, whereas the entropy contribution scales as $N$. Therefore, for large $N$ one can neglect the contribution of the entropy term  in the saddle point calculation. The SPE in Eq.~\eqref{eq:spe2}  now explicitly reads 
\begin{align}
\frac{N x^2}{2}+\frac{\delta \mathcal{E}_I[\rho_N(x)]}{\delta \rho_N(x)}\bigg{|}_{\rho_N(x)=\varrho_N(x)} =\begin{cases}\mu_1 ~~~\text{for}~~~-\ell_N \leq x \leq \ell_N \\
\mu_2 ~~~\text{for}~~~ \ell_N < |x| \leq b_N.
\end{cases}
\label{sad-eq-gen}
\end{align}
In the large $N$ limit the functional $\mathcal{E}_I[\rho_N(x)]$ for a fixed $f$ can be computed (for both the models) following the procedure in \cite{sanaa2019}.
Since the form of the energy functional $\mathcal{E}_I[\rho_N(x)]$ turns out to be different for $1/2\leq f \leq 1$ and $0 < f < 1/2$, we present the derivation of the SPEs for these two cases  separately in the next sections. 

\subsection{For $1/2\leq f\leq 1$:}
\label{sad-eq-f>1/2}
Note in this case $N/2 \leq d \leq N-1$. We first rewrite the interaction part of the energy function $\mathscr{E}(\{x_i\})$ in Eq.~\eqref{E-ocp-lg} as 
\begin{align}
\mathscr{E}_I(\{x_i\})= \frac{1}{2} \left[  \underset{i\ne j}{\sum_{i=1}^N  \sum_{j=1}^{N}} - \sum_{i=1}^{N-d-1}  \sum_{j=i+d}^{j=N}  - \sum_{i=d+2}^N  \sum_{j=1}^{j=i-d} \right]~V(|x_i-x_j|).
\end{align}
Recalling the definition $\rho_N(x)=\frac{1}{N}\sum_{i=1}^N\delta(x-x_i)$, it is easy to convert the different terms into integrals as 
\begin{align}
\mathcal{E}_{I}[\rho_N(x)]=&\frac{N^2}{2}  \int_{-b_N}^{b_N} dx \int_{-b_N}^{b_N} dx'  ~V(x-x')~\rho_N(x) \rho_N(x') \notag \\
&~~~~~~~-\frac{N^2}{2} \int_{-b_N}^{-\ell_N} dx \int_{x+\bar{\Delta}_x}^{b_N} dx' ~V(x-x')~\rho_N(x) ~\rho_N(x')
\label{energy_fgreater2} \\
&~~~~~~~-\frac{N^2}{2}\int_{\ell_N}^{b_N} dx \int_{-b_N}^{x-\Delta_x} dx' ~V(x-x')~\rho_N(x) ~\rho_N(x'), \notag 
\end{align}
where $V(x)$ is specified in Eq.~\eqref{E-ocp-lg}. The distances $\Delta_x$ and $\bar{\Delta}_x$ depend on density profile $\rho_N(x)$ as
\beq
\int_{x-\Delta_x}^{x} \rho_N(x') dx'=f=\int_{x}^{x+\bar{\Delta}_x}\rho_N(x') dx'.
\label{Delta_x}
\eeq
These equations define $\Delta_x$ and $\bar{\Delta}_x$ that represent the distance one has to move starting from $x$ to  find $f$ fraction of particles on the left and right side, respectively. Note these quantities, being addition to function of $x$,  are also functionals of $\rho_N(x)$. In the following, we will refer them as functions of $x$.

Note that, in writing down the interaction energy part $\mathcal{E}_{I}$ of the energy functional, we have ignored possible self-energy contribution~\cite{Dean_PRE_2008,Sandier_arxiv_2013}. The reason behind this is that, in the large $N$ limit the self-energy term remains subdominant compared to other terms in the free energy. Such self energy contribution for ATAC log-gas ($f=1$) was shown to be of order $\sim N \int dx \rho_N(x) \log(\rho_N(x))$ in Ref.~\cite{Dean_PRE_2008}. While writing the expression of $\mathcal{E}_I[\rho_N(x)]$, given a particle at $x$ we are integrating $x'$ over a region including the point $x$. For the log-gas interaction, two particles cannot stay at the same point. Hence the region near point $x$ should be excluded in the integration of $x'$ in the log-gas case. However for $1$dOCP, one does not face any such problem. In order to correct for that, one needs to subtract a self energy contribution which for large $N$ can be estimated as 
\beq
\mathscr{E}_{\text{self}}\sim N^2 \int_{-b_N}^{b_N} dx \rho_N(x) \int_{x-\epsilon_x}^{x+\epsilon_x} dx' \rho_N(x') V(|x-x'|).
\label{E-self}
\eeq
Here $\epsilon_x$ represents the smallest gap between two particles at position $x$, which in the large $N$ limit for a given density profile $\rho_N(x)$ is given by $\epsilon_x \approx \frac{1}{N \rho_N(x)}$~\cite{dyson_JKP_1962}. After a few straightforward algebraic steps, the self-energy term for the log-gas ($V(x)=-\log(|x|)$) simplifies to 
\beq
\mathscr{E}_{\text{self}} \sim N \int_{-b_N}^{b_N} \rho_N(x) \log[\rho_N(x)] dx + \text{constant}. \label{E-self-2}
\eeq
This term, however, is much smaller than the bulk interaction energy term of $\mathcal{O}(f N^2)$ in the large $N$ limit. Hence, one can neglect the self energy contribution in the saddle point calculation for fixed arbitrary value of $f$.

 \noindent To obtain the SPE  \eqref{sad-eq-gen} explicitly, we perform functional derivative of $\mathcal{E}[\rho_N(x)]$ and for that we use the following results 
\begin{align}
\frac{\delta \Delta_{x}}{\delta \rho_N(z)}&=-\frac{1}{\rho_N(x-\Delta_x)} \Theta(z-x+\Delta_x) \Theta(x-z).\label{Delta-rho} \\
\frac{\delta \bar{\Delta}_x}{\delta \rho_N(z)}&=-\frac{1}{\rho_N(x+\bar{\Delta}_x)} \Theta(x+\bar{\Delta}_x-z) \Theta(z-x). \label{bDelta-rho}
\end{align}
Performing the functional derivatives and using the above equations, we finally get the SPEs in all three regimes. The equation in the central regime $-\ell_N \le x \le \ell_N$ is given by 
\begin{align}
N& \frac{x^2}{2}+ N^2~ \int_{-b_N}^{b_N}  V(x'-x)~\varrho_N(x')~ dx' \label{SPE_genk_f>1/2_center} \\
& -\frac{N^2}{2} \Big[ \int_{-b_N}^{-\ell_N}  V(\bar{\Delta}_{x'})~\varrho_N(x')~ dx' + \int_{\ell_N}^{b_N} V(\Delta_{x'})~\varrho_N(x') ~ dx' \Big]  
=\mu_1.
\notag
\end{align} 
The SPEs in the left regime  $-b_N \le x<-\ell_N$ and the right regime $\ell_N<x \le b_N$ are, respectively,  
\begin{align}
&N\frac{x^2}{2} + N^2~  \int_{-b_N}^{b_N}  V(x'-x)~\varrho_N(x')~ dx' -\frac{ N^2}{2}~ \left[2 \int_{x+\bar{\Delta}_x}^{b_N}  V(x'-x)\varrho_N(x')dx' \right. \notag  \\
& ~~~~~~~~~~~~~~~~~\left.+  \int_{-b_N}^{x} V(\bar{\Delta}_{x'}) \varrho_N(x')dx' +\int_{\ell_N}^{x+\bar{\Delta}_x} V(\Delta_{x'})\varrho_N(x')dx' \right] =\mu_2.
\label{SPE_genk_f>1/2_left} 
\end{align}
and
\begin{align}
&N\frac{x^2}{2} + N^2~  \int_{-b_N}^{b_N}  V(x'-x)~\varrho_N(x')~ dx'  -\frac{ N^2}{2}~ \left[2 \int_{-b_N}^{x-\Delta_x}  V(x'-x) \varrho_N(x') dx'  \right. \notag  \\
&~~~~~~~~~~~~~~~~~\left.+  \int_{x}^{b_N} V(\Delta_{x'}) \varrho_N(x') dx' +\int_{x-\Delta_x}^{-\ell_N} V(\bar{\Delta}_{x'}) \varrho_N(x') dx' \right] =\mu_2.
\label{SPE_genk_f>1/2_right} 
\end{align}
Note in the ATAC ($f\to1$) limit, the central part grows and both the edge regions shrink to the edge points of the central part. The terms inside the square brackets of Eqs.~(\ref{SPE_genk_f>1/2_center} - \ref{SPE_genk_f>1/2_right}) go to zero and consequently the SPEs corresponding to the edge parts reduce to the SPE of the central part evaluated at the edge points with $\mu_2=\mu_1$.

\subsection{For $0<f <1/2$}
\label{sad-eq-f<1/2}
In this case $0<d\leq N/2$, we rewrite the microscopic energy function $\mathscr{E}(\{x_i\})$ in Eq.~\eqref{E-ocp-lg} as 
\begin{align}
\mathscr{E}_I(\{x_i\})= \frac{1}{2} \left[\underset{i\ne j}{\sum_{i=1}^d  \sum_{j=1}^{i+d}} + \underset{i\ne j}{\sum_{i=d+1}^{N-d}  \sum_{j=i-d}^{j=i+d}}  + \underset{i\ne j}{\sum_{i=N-d+1}^N  \sum_{j=i-d}^{j=N}}  \right]~V(|x_i-x_j|).
\end{align}
 In terms of the empirical density $\rho_N(x)=\frac{1}{N}\sum_{i=1}^N\delta(x-x_i)$, we convert the double sums in the above equation to double integrals and find
\begin{align}
&\mathcal{E}_{I}[\rho_N(x)]=\frac{N^2}{2} \int_{-b_N}^{-\ell_N} dx \int_{-b_N}^{x+\bar{\Delta}_x} dx' V(x-x')\rho_N(x) \rho_N(x') \notag \\
&~~~~~~~~~~~~~~~~~~+\frac{ N^2}{2}  \int_{-\ell_N}^{\ell_N} dx \int_{x-\Delta_x}^{x+\bar{\Delta}_x} dx' ~V(x-x')~\rho_N(x) ~\rho_N(x') \label{eq:threeparts2} \\
&~~~~~~~~~~~~~~~~~~~ +  \frac{N^2}{2} \int_{\ell_N}^{b_N} dx \int_{x-\Delta_x}^{b_N} dx' V(x-x')\rho_N(x) \rho_N(x'), \notag 
\end{align}
where the distance functions $\Delta_x$ and $\bar{\Delta}_x$ are defined in Eq.~\eqref{Delta_x}.  
In this case also we have neglected  the self energy contribution for $V(x)=-\log(|x|)$.  As can be seen from Eq.~\eqref{E-self-2}, this contribution can be neglected for $0<f<1/2$ also as long as it is kept fixed at a non-zero value (however small it could be) in the $N \to \infty$ limit. 
Inserting the above form for $\mathcal{E}_I[\rho_N(x)]$ in Eq.~\eqref{sad-eq-gen}, one can get the SPEs for $0<f < 1/2$.
The SPE in the left regime $-b_N \le  x < -\ell_N$ is given by 
\begin{align}
\frac{Nx^2}{2}+& N^2\int_{-b_N}^{x+\bar{\Delta}_x}  V(x'-x)~\varrho_N(x')~ dx'  \label{SPE_genk_f<1/2_left} \\
&-\frac{N^2}{2}\left[ \int_{-b_N}^{x} V(\bar{\Delta}_{x'})\varrho_N(x')~dx'+ \int_{-\ell_N}^{x+\bar{\Delta}_x}   V(\Delta_{x'}) \varrho_N(x')~dx' \right]  =\mu_2,\notag 
\end{align}
and  in the right regime $\ell_N < x \le  b_N$ we get
\begin{align}
\frac{Nx^2}{2}&+N^2 \int_{x-\Delta_x}^{b_N} V(x'-x)~\varrho_N(x')~dx' \label{SPE_genk_f<1/2_right}\\
&-\frac{N^2}{2}\left[ \int_{x}^{b_N} V(\Delta_{x'})~\varrho_N(x')~dx'+ \int_{x-\Delta_x}^{\ell_N}   V(\bar{\Delta}_{x'})~\varrho_N(x')~dx'\right] =\mu_2. \notag 
\end{align}
In the central regime $-\ell_N \le x \le \ell_N$ we get
\begin{align}
\frac{Nx^2}{2}+&N^2 \int_{x-\Delta_x}^{x+\bar{\Delta}_x} V(x'-x) ~\varrho_N(x') ~dx'\notag \\
&-\frac{N^2}{2}\left[ \int_{x-\Delta_x}^{x}   V(\bar{\Delta}_{x'}) ~\varrho_N(x')~dx'+ \int_{x}^{x+\bar{\Delta}_x} V(\Delta_{x'}) ~\varrho_N(x')~dx' \right] \notag \\
&+ \frac{N^2}{2}
\int_{-b_N}^{-\ell_N}  \left[ V(x'-x) -  V(\bar{\Delta}_{x'}) \right ] \Theta(x'+\bar{\Delta}_{x'}-x)  ~\varrho_N(x')~dx' \label{SPE_genk_f<1/2_center} \\
&+\frac{N^2}{2} \int_{\ell_N}^{b_N} \left [ V(x'-x) -  V(\Delta_{x'}) \right ] \Theta(x-x'+\Delta_{x'}) ~\varrho_N(x')~dx'=\mu_1. \notag 
\end{align}

 \section{Explicit solution of the saddle point equation for FR $1$dOCP} 
 \label{Solution_sp_1docp}
In this Appendix, we provide details on solving the SPEs of the $1$dOCP model for arbitrary $f \in (0,1]$. We present the solutions in different regions for $1/2 \le f \leq 1$ and $0< f < 1/2$ separately. 

  \subsection{The saddle point calculation for $1/2 \le f \leq 1$:}
  \label{sp-sol-fr1docp-f>1/2}
 The SPE for this case in the central region $[-\ell_N,\ell_N]$ is given in Eq.~\eqref{SPE_genk_f>1/2_center} whereas the SPEs on the left part $[-b_N,-l_N)$ and right  part $(l_N,b_N]$ are given in Eqs.\eqref{SPE_genk_f>1/2_left} and \eqref{SPE_genk_f>1/2_right} respectively with $V(x)=-|x|$.
 From numerical results in Fig.~\ref{largef_kmone_figure}, we observe that both the length scales $\ell_N$ and $b_N$ are $ \sim \mathcal{O}(N)$ and the equilibrium density indeed has the scaling form $\varrho_N(x) = \frac{1}{N}\tilde{\varrho}_f\left( \frac{x}{N}\right)$. Hence it is natural to work in the scaling variable $y=x/N$ and with the scaled density profile $\tilde{\varrho}_f(y)$. The SPE in the central regime can be written as
 \begin{align}
 \frac{y^2}{2} & -  \int_{-\tilde{b}}^{\tilde{b}} |y'-y|~ \tilde{\varrho}_f(y') ~dy' \notag \\
&+\frac{1}{2} \Big[ \int_{-\tilde{b}}^{-\tilde{\ell}} \bar{\delta}_{y'} ~\tilde{\varrho}_f(y')~dy' + \int_{\tilde{\ell}}^{\tilde{b}} \delta_{y'}  \tilde{\varrho}_f(y') ~dy' \Big]=\tilde{\mu}_1,
\label{sp_1docp_flarge_central_scaled2}
\end{align}
where $y=x/N$, $\tilde{b}=b_N/N$, $\tilde{\ell}=\ell_N/N$ and $\tilde{\mu}_1=\mu_1/N^3$. The SPE on the left edge part can be written  
in terms of the scaled quantities as
\begin{align}
&\frac{y^2}{2} -  \int_{-\tilde{b}}^{\tilde{b}} |y'-y|~ \tilde{\varrho}_f(y')~dy' \notag  \\
          & +\frac{1}{2}~ \left(2 \int_{y+\bar{\delta}_y}^{\tilde{b}}  |y'-y| \tilde{\varrho}_f(y')~dy' +  \int_{-\tilde{b}}^{y} \bar{\delta}_{y'} \tilde{\varrho}_f(y')~ dy' +\int_{\tilde{\ell}}^{y+\bar{\delta}_y} \delta_{y'} \tilde{\varrho}_f(y')~ dy' \right) =\tilde{\mu}_2.
\label{eq:sp_1docp_flarge_edge_scaled}
\end{align}
where $\tilde{\mu}_2=\mu_2/N^3$. A similar equation can be written for the right edge part as well. 
Note that in writing the above equations we have assumed that the functions $\Delta_x$ and $\bar{\Delta}_x$ [defined in Eq.~\eqref{Delta_x}] have scaling forms $\bar{\Delta}_{x=Ny}=N\bar{\delta}_y$ and $\Delta_{x=Ny}=N\delta_y$, respectively.

From Fig.~\ref{largef_kmone_figure} we further observe that the scaled density profile is piece-wise uniform. This leads us to make the following ansatz for the scaled density  profile as 
 \beq
 \tilde{\varrho}_f(y)=\begin{cases}
     \tilde{\varrho}_{\text{edge}} ~\text{for}~ -\tilde{b} \le y<-\tilde{\ell} \\
     \tilde{\varrho}_{\text{mid}} ~~\text{for}~ -\tilde{\ell} \le y \le \tilde{\ell} \\
     \tilde{\varrho}_{\text{edge}} ~~\text{for}~ ~~~\tilde{\ell}<y \le \tilde{b}.    
 \end{cases}
  \label{density_profile_scaled_a}
 \eeq
Our next task is to express the SPEs in terms of the above ansatz for $\tilde{\varrho}_f(y)$ and solve for the six  unknown constants $\tilde{\varrho}_{\rm mid}$, $\tilde{\varrho}_{\rm edge}$, $\tilde{\ell}$, $\tilde{b}$, $\tilde{\mu}_1$ and $\tilde{\mu}_2$. To proceed further, we first need to determine the $y$ dependence of the functions $\delta_y$ and $\bar{\delta}_y$. Recall, these functions determine the lengths, starting from $y$, one needs to cover on the left or right to find $f$ fraction of particles. They are defined as 
 \beq
\int_{y-\delta_y}^{y} \tilde{\varrho}_f(y') dy'=\int_{y}^{y+\bar{\delta}_y}\tilde{\varrho}_f(y') dy'=f.
\label{density_constraint-app}
\eeq
Note these equations are scaled version of Eq.~\eqref{Delta_x}.
For the particle at $y'=-\tilde{b}$, the distance $\bar{\delta}_{y'}$ is $(\tilde{\ell}+\tilde{b})$ since exactly $f$ fraction of total particles are present inside $[-\tilde{b},\tilde{\ell}]$ and the remaining $(1-f)$ fraction of particles stay inside $(\tilde{\ell},\tilde{b}]$. As the equilibrium density is equal to $\tilde{\varrho}_{\text{edge}}$ both in the left part $[-\tilde{b},-\tilde{\ell})$ and in the right part $(\tilde{\ell},\tilde{b}]$, the distance $\bar{\delta}_{y'}$ for any $y' \in [-\tilde{b},-\tilde{\ell}]$ is exactly  equal to $(\tilde{\ell}+\tilde{b})$. Similar arguments hold for $\delta_{y'}$ when $y' \in [\tilde{\ell},\tilde{b}]$.  As a result $\delta_{y'}$ for any $y' \in [\tilde{\ell},\tilde{b}]$ is also equal to $(\tilde{\ell}+\tilde{b})$. The function $\delta_y$ is defined only over the domain $[\tilde{\ell},\tilde{b}]$ and $\bar{\delta}_y$ over $[-\tilde{b},-\tilde{\ell}]$. They have the following explicit  forms
\begin{align}
\begin{split}
\bar{\delta}_y&=(\tilde{b}+\tilde{\ell}) ~~~~\text{for}~ -\tilde{b} \le y \le -\tilde{\ell} \\
\delta_y&=(\tilde{b}+\tilde{\ell}) ~~~~\text{for}~~~~~ \tilde{\ell} \le y \le \tilde{b}. 
\end{split}
\label{eq:delta_functional_form}
\end{align}
Inserting the form of $\tilde{\varrho}_f(y)$ from Eq.~\eqref{density_profile_scaled_a} and the forms of $\delta_y$ and $\bar{\delta}_y$ from Eq.~\eqref{eq:delta_functional_form} in the SPE~\eqref{sp_1docp_flarge_central_scaled2}, and performing the integrals one finds  
\beq
\Big(\frac{1}{2} - \tilde{\varrho}_{\text{mid}} \Big) y^2 - \tilde{\varrho}_{\text{mid}}\tilde{\ell}^2 =\tilde{\mu}_1,~~\text{for}~~-\tilde{\ell} \le  y \le  \tilde{\ell}.
\label{eq:fgreater_central_final_form}
\eeq
Since this equation is valid for all $-\tilde{\ell}\le y \le \tilde{\ell}$, the coefficients of all the powers of $y$ must vanish individually,  which in turn implies 
\begin{align}
\tilde{\varrho}_{\text{mid}}=\frac{1}{2},~\tilde{\ell}=(2f-1) ~\text{and}~ \tilde{\mu}_1=-\tilde{\varrho}_{\text{mid}} \tilde{\ell}^2.
 \label{eq:findings_1}
\end{align}

We now focus on solving the SPEs on the edges. We start with the SPE on the left edge part given in Eq.~\eqref{eq:sp_1docp_flarge_edge_scaled}, where once again we 
 insert the ansatz for $\tilde{\varrho}_f(y)$ from Eq.~\eqref{density_profile_scaled_a} and the forms of  $\delta_y$ and $\bar{\delta}_y$ from  Eq.~\eqref{eq:delta_functional_form}. After performing the integrals we get  
\beq
 \Big(1- \tilde{\varrho}_{\text{edge}} \Big) \frac{y^2}{2}+2\tilde{\ell} \left(\tilde{\varrho}_{\text{mid}} -\frac{\tilde{\varrho}_{\text{edge}}}{2} \right)y- \tilde{\varrho}_{\text{edge}} \frac{\tilde{\ell}^2}{2}=\tilde{\mu}_2.
\label{eq:sp_fg_final_edge}
\eeq
This equation is valid for any $y \in [-\tilde{b}, -\tilde{\ell})$. Consequently,  the coefficients of all the powers of $y$ must vanish individually which provides 
\begin{align}
 &\tilde{\varrho}_{\text{edge}}=1,~ \tilde{\varrho}_{\text{mid}}=\frac{\tilde{\varrho}_{\text{edge}}}{2}=\frac{1}{2} ~\text{and}~ \tilde{\mu}_2=- \tilde{\varrho}_{\text{mid}} \tilde{\ell}^2.
 \label{eq:findings_2}
\end{align}
From Eq.~\eqref{eq:findings_1} and Eq.~\eqref{eq:findings_2} we notice $\tilde{\mu}_1=\tilde{\mu}_2$. One can perform a similar calculation for the right part and find exactly the same expressions for the constants as in Eq.~\eqref{eq:findings_2}.
Combining Eqs.~\eqref{eq:findings_1} and \eqref{eq:findings_2} we get the values of all the constants appearing in Eq.~\eqref{density_profile_scaled_a}
\begin{align}
 &\tilde{\varrho}_{\text{edge}}=1,~~\tilde{\varrho}_{\text{mid}}=\frac{1}{2}, ~~\tilde{b}=f,~~\tilde{\ell}=(2f-1), 
 \label{eq:findings_all}
\end{align}
and thus completely specify the scaled density profile $\tilde{\varrho}_f(y)$ defined in Eq.~\eqref{density_profile_scaled_a}.

\subsection{The saddle point calculation for $0 < f < 1/2$:}
 \label{sp-sol-fr1docp-f<1/2}
In this case, there are $(1-2f)$ fraction of particles in the central region and $f$ fractions of particles on either side of it in equilibrium.
The SPEs in all the three regions, centre, left and right, can be read off  from Eqs.~(\ref{SPE_genk_f<1/2_left}- \ref{SPE_genk_f<1/2_center} ) with $V(x)=-|x|$. One can rewrite these equations in terms of the scaled variable $y=x/N$ which will turn out to be convenient as shown below. 
As discussed in Sec.~\ref{saddle-1docp-text-f>1/2} (see Eq.~\eqref{eq:density_k_mone_scaled_text}), 
 we expect that for $0<f<1/2$  the saddle point (scaled) density profile should have the following form:
\beq
\tilde{\varrho}_f(y)=(1-2f) \delta(y)+\tilde{\varrho}_{\text{edge}} \mathbbm{1}_{[-\tilde{b},\tilde{b}]}(y).
\label{eq:density_k_mone_scaled}
\eeq
Our task is to find the density value $\tilde{\varrho}_{\rm edge}$ and the size $\tilde{b}$ of the support. For that we 
first rewrite the SPE  from Eq.~\eqref{SPE_genk_f<1/2_left} in terms of scaling variable $y=x/N$ on the left part $-\tilde{b} \le  y < 0_-$: 
\begin{align}
\begin{split}
&\frac{y^2}{2} \underbrace{-  \int_{-\tilde{b}}^{0_-} |y-y'|  \tilde{\varrho}_f(y') dy'}_{T_2}~~
\underbrace{- \int_{0_-}^{y+\bar{\delta}_y} (y'-y) \tilde{\varrho}_f(y') dy'}_{T_3}   \\
& ~~~~~
 +\underbrace{\frac{1}{2} \int_{-\tilde{b}}^{y} \bar{\delta}_{y'} \tilde{\varrho}_f(y') dy'}_{T_4}+\underbrace{ \frac{1}{2} \int_{0_-}^{y+\bar{\delta}_y} \delta_{y'} \tilde{\varrho}_f(y') dy'}_{T_5}~~ -\tilde{\mu}_2 =0.
 \end{split}
\label{eq:se_fless_edge_kmone_scaled}
\end{align}
Note, while writing the above saddle point equation we have assumed that $\tilde{\ell}=\ell_N/N \to 0$ as $N \to \infty$. We will later show that $\ell_N \sim \sqrt{\log N}$ [see Eq.~\eqref{l-sim-sq-logN}] and for the time being we assume $\tilde{\ell} \to 0$  and proceed. Also note, we have assumed that the functions 
$\bar{\Delta}_x$ and $\Delta_x$ have the following scaling $\bar{\Delta}_{x=yN}=N\bar{\delta}_y$ and $\Delta_{x=yN}=N\delta_y$. Explicit forms of these functions will be determined later. 

Next task is to insert  the ansatz from Eq.~\eqref{eq:density_k_mone_scaled} for the scaled density profile  in Eq.~\eqref{eq:se_fless_edge_kmone_scaled}, and evaluate different integrals in this equation. The second term in the L.H.S. of Eq.~\eqref{eq:se_fless_edge_kmone_scaled} can be evaluated easily and one finds 
\begin{align}
  T_2=-  \int_{-\tilde{b}}^{0_-} |y-y'|  \tilde{\varrho}_f(y') dy'
  =-\tilde{\varrho}_{\text{edge}} \Big[y^2+y \tilde{b}+\frac{\tilde{b}^2}{2}\Big].
  \label{eq:second_term}
\end{align}
To evaluate the other terms we first need to find the $y$ dependence of $\delta_y$ and $\bar{\delta}_y$. It turns out that the $y$ dependence of  $\bar{\delta}_y$ and $\delta_y$  are different for $1/3\leq f < 1/2$ and $0<f < 1/3$. This happens because for $1/3\leq f <1/2$, the fraction of particles $(1-2f)$ in the central region is smaller than the fraction $f$ of particles on either edge parts, whereas for $0<f<1/3$, the fraction in the central part is larger than $f$.  Hence, from this point one requires to consider the two cases separately. 

\subsubsection{$1/3 \le f<1/2$:}
\label{1docp-sd-1/3<f<1/2}
In order to determine $y$ dependence of $\delta_y$ and $\bar{\delta}_y$, we first need to investigate the $x$ dependence of $\Delta_x$ and $\bar{\Delta}_x$. As shown in Fig.~\ref{smallf_kmone_figure}(b), the density profile in the central part can be described by a Gaussian distribution function of the unscaled variable $x$. 
On the other hand from Fig.~\ref{smallf_kmone_figure}(a), we observe that the density profile outside the central part is constant and to be consistent with the ansatz for the scaled profile in Eq.~\eqref{eq:density_k_mone_scaled}, the value of the constant should be $\frac{\tilde{\varrho}_{\rm edge}}{N}$. It is easy to see that for $x=-b_N$ one has to move to point $x=-\ell_N$ {\it i.e.} by a distance $\bar{\Delta}_{-b_N}=b_N-\ell_N$ on the right to find $f$ fraction of particles. Now if we start from a point $x$ slightly above $x=-b_N$, one has to reach a point $\bar{X}(x)=x+\bar{\Delta}_x$ inside the central region $[-\ell_N,\ell_N]$ in order to find $f$ fraction of particles on the right. As the point $x$ moves towards right starting from $x=-b_N$, the image point $\bar{X}(x)$ also moves towards right and for $1/3 \le f<1/2$ there will be a point $x=-x_f$ for which the image point will touch the right boundary of the central region {\it i.e.} $\bar{X}(-x_f)=\ell_N$. It is easy to see that for $-b_N\le x<-x_f$, $\bar{X}(x)$ can be obtained by solving 
\beq
\int_{-\ell_N}^{\bar{X}(x)} \rho_{\rm central}(x') dx' = (b_N+x)\frac{\tilde{\varrho}_{\rm edge}}{N}, \label{barh_x}
\eeq
where $\rho_{\rm central}(x)$ is density profile in the central part described by a Gaussian function as shown in Fig.~\ref{smallf_kmone_figure}(b) [see Eq.~\eqref{rho(x)-ansatz)} and Eq.~\eqref{eq:density_central_kmone_text}]. Since the central region contains $(1-2f)$ fraction of particles, $|\bar{X}(x)|<\ell_N$ for $-b_N \le x<-x_f$ and $\bar{X}(-x_f)=\ell_N$. Hence, one finds the following expression for $x_f$
\beq 
x_f = \frac{(3f-1)N}{\tilde{\varrho}_{\rm edge}}+\ell_N. \label{x_f}
\eeq
Now as the point $x$ moves further right of $-x_f$, the corresponding image point $\bar{X}(x)$ enters the right edge region where the density profile is again uniform with value $\frac{\tilde{\varrho}_{\rm edge}}{N}$. So it becomes straightforward to show that $\bar{\Delta}_{x} = x_f +\ell_N$ {\it i.e.} independent of $x$ for $-x_f \le x<-\ell_N$. Now as $x$ enters the region $[-\ell_N,\ell_N]$, we expect $\bar{\Delta}_x$ should again start depending on $x$ and should change rapidly by large amount over a small distance $2l_N$ such that $\bar{X}(\ell_N)=b_N$,  the rightmost point of the support of the full density profile. Performing a similar estimate as done for the range $-b_N \le x<-x_f$, one can show that $\bar{\Delta}_x = x_f+ \frac{N}{\tilde{\varrho}_{\rm edge}} \int_{-\ell}^x \rho_{\rm central}(x')dx'-x$ for $-\ell_N \leq x \leq \ell_N$. Collecting the functional forms of $\bar{\Delta}_x$ from different regions we, for $1/3 \leq f<1/2$, have
\begin{align}
 \bar{\Delta}_x = 
 \begin{cases}
     \bar{X}(x)-x ~&~~\text{for}~~-b_N\le x\le -x_f \\
     x_f+\ell_N ~ & ~~\text{for}~~-x_f \le  x \le -\ell_N \\
     x_f + \frac{N}{\tilde{\varrho}_{\rm edge}} \int_{-\ell_N}^x \rho_{\rm central}(x')dx' - x ~& ~~\text{for}~~-\ell_N \le x \le \ell_N,
 \end{cases}
 \label{bar_Delta_x}
\end{align}
where $x_f$ is given in Eq.~\eqref{x_f} and $\bar{X}(x)$ is determined from the solution of Eq.~\eqref{barh_x}. In the large $N$ limit the length function $\bar{\Delta}_x$ can be written in terms of the scaling variable $y=x/N$ as $\bar{\Delta}_{x=Ny}= N\bar{\delta}_y$ where  
\begin{align}
 \bar{\delta}_y = 
 \begin{cases}
     -y ~&~~\text{for}~~-\tilde{b}\le y\le -\tilde{x}_f \\
     ~\tilde{x}_f ~ & ~~\text{for}~~-\tilde{x}_f \le y \le 0_- \\
     ~~\tilde{b}  ~& ~~\text{for}~~~~~y=0_+,
 \end{cases}
 \label{eq:delta_tilde_bar}
\end{align}
with 
\beq 
\tilde{x}_f = \frac{x_f}{N}=
\frac{(3f-1)}{\tilde{\varrho}_{\rm edge}}~~\text{for}~\frac{1}{3} \leq f < \frac{1}{2}. \label{tx_f}
\eeq
One can perform a similar analysis to find the $y$ dependence of $\delta_y$. Using the fact that the saddle point density profile should be symmetric about the origin (centre of the trap), one can write 
\begin{align}
 \delta_y = 
 \begin{cases}
     y ~&~~\text{for}~~\tilde{x}_f \le y \le \tilde{b} \\
     \tilde{x}_f ~ & ~~\text{for}~~0_+ \le y \le \tilde{x}_f\\
     \tilde{b}  ~& ~~\text{for}~~~~~y=0_-.
 \end{cases}
 \label{eq:delta_tilde}
\end{align}
Now we are in a position to evaluate the $3$rd, $4$th and $5$th terms on the L.H.S. of Eq.~\eqref{eq:se_fless_edge_kmone_scaled}. We now insert the ansatz for the density profile from Eq.~\eqref{eq:density_k_mone_scaled} and, $\bar{\delta}_y$ and ${\delta}_y$ from Eqs.~(\ref{eq:delta_tilde_bar},~\ref{eq:delta_tilde}), respectively, in the expressions of these terms and perform the integrals. Since $y$ dependence of $\bar{\delta}_y$ is different in different ranges of $y$, one needs to evaluate the integrals for $y$ falling in different ranges separately. We first focus on the third term $T_3$ for $-\tilde{b} \leq y \leq -\tilde{x}_f$, in Eq.~\eqref{eq:se_fless_edge_kmone_scaled}. To evaluate this term, we first note that for $-\tilde{b} \leq y \leq -\tilde{x}_f$, the image point $\bar{Y}(y)=y+\bar{\delta}_y$ satisfies $0_-<\bar{Y}(y)<0_+$ (which in terms of unscaled variable reads $-\ell_N < \bar{X}(x) <\ell_N$. Hence we put $y'=0$ in the integrand and approximate $T_3$ as 
$T_3 \approx y~ \int_{0_-}^{y+\bar{\delta}_y}  \tilde{\varrho}_f(y') dy' $.  
Note that, the integral $\int_{0_-}^{y+\bar{\delta}_y}  \tilde{\varrho}_f(y') dy' $ represents exactly the same fraction of particles inside the region $[-\tilde{b},y]$. This is because one has to acquire same fraction of particles that one looses when one moves from $-\tilde{b}$ to $y~( <-\tilde{x}_f$) so that one finds $f$ fraction of particles between $y$ to $y+\bar{\delta}_y$. For $ -\tilde{x}_f \leq y < 0_-$, it is straightforward to evaluate the integral $T_3$. One can evaluate the integral $T_4$  in the two different ranges of $y$ also separately and once again straightforwardly. One finally has  
\begin{align}
T_3&=
\begin{cases}
y(\tilde{b}+y)\tilde{\varrho}_{\text edge}  &~\text{for}~~-\tilde{b}\leq y\leq -\tilde{x}_f \\
  y (1-2f) - \tilde{\varrho}_{\text{edge}} \Big[ \frac{(y+\tilde{x}_f)^2}{2}-y(y+\tilde{x}_f)\Big] &~\text{for}~~-\tilde{x}_f \leq y <0_-,
\end{cases}
\label{eq:third_term} \\ 
 T_4&=
 \begin{cases}
  \frac{\tilde{\varrho}_{\text{edge}}}{4} \big(\tilde{b}^2-y^2\big) &~~~~~~~~~~~\text{for}~~
  -\tilde{b}\leq y\leq -\tilde{x}_f \\
 \frac{\tilde{\varrho}_{\text{edge}}}{4} \big(\tilde{b}^2-\tilde{x}^2_f\big) +  \frac{\tilde{x}_f \tilde{\varrho}_{\text{edge}}}{2}\big(y+\tilde{x}_f\big), &~~~~~~~~~~~\text{for}~~ 
 -\tilde{x}_f \leq y <0_-,
 \end{cases}
 \label{eq:fourth_term}
\end{align}
To compute the fifth term $T_5$ in Eq.~\eqref{eq:se_fless_edge_kmone_scaled} one needs to be careful and we discuss the computation for the two ranges $-\tilde{b}  \le  y\leq -\tilde{x}_f$ and $-\tilde{x}_f \leq y <0_-$ separately.  We first focus on the first regime, in which the integral is 
\begin{align}
T_5= \frac{1}{2}\int_{0_{-}}^{\bar{Y}(y)}  \delta_{y'} \tilde{\varrho}_f(y')dy', \label{T_5_sc_f>1/3-1}
\end{align}
 where $\bar{Y}(y)=y+\bar{\delta}_y < 0_+$. Although the range of the integral $[0_-,\bar{Y}(y)]$ is very small (in fact infinitesimal), the integral would produce a finite contributions because the (scaled) density profile in this range is a delta function and the function $\bar{\delta}_y$ changes rapidly by a large amount which, in fact, appears to be a discontinuous change in the $N \to \infty$ limit. In order to find the  contribution of the integral we rewrite the integral $T_5$ in terms of the unscaled variables as  
\begin{align}
T_5 &= \frac{1}{2N}\int_{-\ell_N}^{\bar{X}(x)}  \Delta_{x'} \rho_{\rm central}(x')dx'.
\label{eq:T_5_usc_f>1/3-1}
\end{align}
Recall that for large $N$, there are $(1-2f)N$ particles in the central part out of which $n = \tilde{\varrho}_{\text{edge}} (b_N+x)$ particles stay inside $[-\ell_N,\bar{X}(x)]$. In terms of the microscopic positions of these particles, the integral in Eq.~\eqref{eq:T_5_usc_f>1/3-1} can be written as 
\begin{align}
  T_5 &= \frac{1}{2N}\int_{-\ell_N}^{\bar{X}(x)}  \Delta_{x'} \frac{1}{N} \left \langle\sum_{i=1}^{N} \delta(x'-x_i) \right \rangle_{\rm eq} dx'= \frac{1}{2N^2}  \sum_{i=fN+1}^{fN+n} \left \langle\Delta_{x_i} \right \rangle_{\rm eq}.
  \label{eq:T_5_usc_f>1/3-1-dis}
\end{align}
It is easy to see that for the left-most particle inside the central region $[-\ell_N,\ell_N]$ labeled by $i=fN+1$, the distance $\langle \Delta_{x_i}\rangle \approx b_N$ since exactly $f$ fraction of particles are present inside the left edge region $[-b_N,-\ell_N)$. For the next particle labeled by $i=fN+2$, the distance $\langle \Delta_{x_i}\rangle \approx (b_N-\bar{d})$ with $\bar{d}=1/(N\rho_{\rm edge})=\frac{1}{\tilde{\varrho}_{\rm edge}}$. Since the density inside $[-b_N,-\ell_N)$ is uniform $\rho_{\rm edge}$, the equilibrium separation between two consecutive particles inside $[-b_N,-\ell_N)$ is $\bar{d}$. Hence for a particle labeled by $i \in [fN+1,fN+n]$ with $n=\tilde{\varrho}_{\text{edge}} (b_N+x)$, the distance $\langle \Delta_{x_i} \rangle_{\rm {eq}} \approx b_N-(i-fN-1) \bar{d}$. Putting this form in Eq.~\eqref{eq:T_5_usc_f>1/3-1-dis} and performing the summation one finds
\beq
T_5 \approx \frac{1}{2N^2}\sum_{j=0}^{n-1}(b- j~ \bar{d})
\approx \frac{1}{4} \tilde{\varrho}_{\text{edge}} (\tilde{b}^2-y^2) 
~~\text{for}~~-\tilde{b}\leq y \leq -\tilde{x}_f.
\label{eq:fifth_term_fless}
\eeq
We now evaluate the $5^{\rm th}$ integral $T_5$ in Eq.~\eqref{eq:se_fless_edge_kmone_scaled} for $-\tilde{x}_f \leq y <0_-$. For $y$ in this range, one finds $0_+ \leq \bar{Y}(y) <  \tilde{x}_f$. Hence we decompose the integral as 
\begin{align}
T_5&
=\underbrace{\frac{1}{2}\int_{0_{-}}^{0_{+}}  \delta_{y'} \tilde{\varrho}_f(y')dy'}_{T_{51}} + \underbrace{\frac{1}{2}\int_{0_{+}}^{\bar{Y}(y)}  \delta_{y'} \tilde{\varrho}_f(y')dy'}_{T_{52}}.
\label{eq:fifth_term} 
\end{align}
Note that the first term in the above equation, denoted by $T_{51}$, has the same form as the integral in Eq.~\eqref{T_5_sc_f>1/3-1} except that now the upper limit is $0_+$. As done for the integral  in Eq.~\eqref{T_5_sc_f>1/3-1},  the integral in Eq.~\eqref{eq:fifth_term} once again can be expressed in terms of the unscaled variables as 
\begin{align}
T_{51} &= \frac{1}{2N}\int_{-\ell_N}^{\ell_N}  \Delta_{x'} \rho(x')dx'= \frac{1}{2N}\int_{-\ell_N}^{\ell_N}  \Delta_{x'} \rho_{\text{central}}(x')dx'.
  \label{eq:t_5_usc}
\end{align}
which can again be evaluated as 
\begin{align}
T_{51} &= \frac{1}{2N}\int_{-\ell_N}^{\ell_N}  \Delta_{x'} \frac{1}{N} \left \langle\sum_{i=1}^{N} \delta(x'-x_i) \right \rangle_{\rm eq} dx'= \frac{1}{2N^2}  \sum_{i=fN+1}^{(1-f)N} \left \langle\Delta_{x_i} \right \rangle_{\rm eq},
  \label{eq:t_5_usc-dis}
\end{align}
where  for a particle labeled by $i \in [fN+1,(1-f)N]$, the distance $\langle \Delta_{x_i} \rangle \approx b_N-(i-fN-1)\bar{d}$. Putting this form in Eq.~\eqref{eq:t_5_usc-dis} and performing the summation one finds
\begin{align}
T_{51}\approx \frac{1}{2N^2}\sum_{j=0}^{(1-2f)N-1}(b- j~\bar{d}) 
=  \frac{1}{2} (1-2f) \Big[\tilde{b}-\frac{\bar{d}}{2}(1-2f) \Big].
\label{eq:fifth_term_1}
\end{align}
We now have to evaluate the remaining term $T_{52}$ on the R.H.S. of Eq~\eqref{eq:fifth_term}. We first note that for  $-\tilde{x}_f \leq y <0_-$, the image point lies within $0_+ \leq \bar{Y}(y) < \tilde{x}_f$ and from Eq.~\eqref{eq:delta_tilde} we observe that $\delta_{y}=\tilde{x}_f$. Putting this form of $\delta_y$ and $\tilde{\varrho}_f(y)=\tilde{\varrho}_{\rm edge}$ inside the integral $T_{52}$ in Eq.~\eqref{eq:fifth_term} and performing the integral, we get 
\beq
T_{52}=\frac{1}{2}\tilde{\varrho}_{\text{edge}} \tilde{x}_f (y+\tilde{x}_f).   \label{eq:fifth_term_2}
\eeq
Hence, collecting the results from Eqs.~\eqref{eq:fifth_term_fless}, \eqref{eq:fifth_term_1} and \eqref{eq:fifth_term_2}, we finally have 
\begin{align}
T_5=
\begin{cases}
 \frac{1}{4} \tilde{\varrho}_{\text{edge}} (y+\tilde{b}) (\tilde{b}-y) 
 &\text{for}~-\tilde{b}\leq y \leq -\tilde{x}_f, \\
  \frac{1}{2} (1-2f) \Big[\tilde{b}-\frac{(1-2f)}{2\tilde{\varrho}_{\rm edge}} \Big] + \frac{1}{2}\tilde{\varrho}_{\text{edge}} \tilde{x}_f (y+\tilde{x}_f), & \text{for}~-\tilde{x}_f \le  y <0_-.
\end{cases}
\label{T_5}
\end{align}
Now we have values of all the integrals in Eq.~\eqref{eq:se_fless_edge_kmone_scaled}. Putting the expressions of the integrals  from Eqs.~\eqref{eq:second_term}, \eqref{eq:third_term},\eqref{eq:fourth_term},\eqref{T_5}, we rewrite Eq.~\eqref{eq:se_fless_edge_kmone_scaled} as 
\begin{eqnarray}
\Big(&1- \tilde{\varrho}_{\text{edge}} \Big ) \frac{y^2}{2} \approx \tilde{\mu}_2, &~\text{for}~-\tilde{b}\leq y\leq -\tilde{x}_f, \notag \\
(&1- \tilde{\varrho}_{\text{edge}}) \frac{y^2}{2} +   \tilde{\varrho}_{\text{edge}} \Big( \frac{f}{\tilde{\varrho}_{\text{edge}}}-\tilde{b} \Big)  y - \frac{ \tilde{\varrho}_{\text{edge}}}{4}&\Big[\tilde{b}^2-\left ( \frac{3f-1}{\tilde{\varrho}_{\text{edge}}}\right )^2\Big] \label{eq:simplification_2} \\
&~~~+\frac{1}{2} (1-2f) \Big[\tilde{b}-\frac{(1-2f)}{2\tilde{\varrho}_{\rm edge}} \Big] \approx \tilde{\mu}_2, &~ \text{for}~-\tilde{x}_f\leq y < 0_-.  \notag 
\end{eqnarray}
Equating coefficients of different powers of $y$ to zero individually in both the regions of $y$ and solving the resulting equations, we find 
\beq
\tilde{\varrho}_{\text{edge}}=1,~~\tilde{b}=Jf ~~\text{and}~~ \tilde{\mu}_2=0,
\label{eq:solution_finbetween}
\eeq
as announced in Eq.~\eqref{consts-f>1/3}. 
 Following a similar procedure, it is straightforward to check that one would get  the same solution for $\tilde{\varrho}_f(y)$ if one solves the saddle point equation in the right part $(0_+<y \leq  \tilde{b}$). We thus get the full solution of the scaled density profile $\tilde{\varrho}_f(y)$ given in Eq.~\eqref{eq:density_k_mone_scaled}.

\subsubsection{$0<f<1/3$:}
\label{1docp-sd-0<f<1/3}
\noindent Having discussed the saddle point calculation for $1/3 \leq f<1/2$, we now move to the $0<f<1/3$ case. To find  $\bar{\Delta}_x$ one can follow a similar analysis as done  in Eq.~\eqref{bar_Delta_x} and get 
\begin{align}
 \bar{\Delta}_x = 
 \begin{cases}
     \bar{X}(x)-x ~&~~\text{for}~~-b_N\le x\le x^*_f \\
     \ell_N + \frac{N}{\tilde{\varrho}_{\rm edge}} \int_{x^*_f}^x \rho_{\rm central}(x')dx' - x ~& ~~\text{for}~~~~x^*_f< x \le \ell_N,
 \end{cases}
 \label{bar_Delta_x_f<1/3}
\end{align}
where $x_f^*$ is the special position such that it's image point $\bar{X}(x^*_f)$ on right is equal to $\ell_N$. It is easy to see that $\bar{X}(x)$ satisfies
\begin{align}
\int_{\max(x,-\ell_N)}^{\bar{X}(x)}\rho_{\rm central}(x')dx' = \frac{[b_N+\min(x,-\ell_N)]\tilde{\varrho}_{\rm edge}}{N}. 
\label{barX-f<1/3}
\end{align}
From the above equation, one finds that $x^*_f$ should satisfy 
\begin{align}
\int_0^{x^*_f}\rho_{\rm central}(x')dx' = \frac{1-4f}{2}, 
\label{x^*_f}
\end{align}
from which it is easy to see that $|x^*_f|<\ell_N$ for $0<f<1/3$. 
Also note that for $-b_N\le x \leq x^*_f$, the image point  $\bar{X}(x)$ falls in the range $[-\ell_N,\ell_N]$ {\it i.e.} $|\bar{X}(x)| \leq \ell_N$. Additionally, using the definition of $x^*_f$ from Eq.~\eqref{x^*_f} in Eq.~\eqref{bar_Delta_x_f<1/3}, one can see that  $ \bar{\Delta}_x  \to (b_N-\ell_N)$ as $x \to \ell_N$. Hence, it is easy to see that for large $N$, $\bar{\Delta}_{x}$ behaves as  $\bar{\Delta}_{x} = N \bar{\delta}_{x/N}$ where 
\begin{align}
\bar{\delta}_{y} = 
\begin{cases}
-y ~~&~~\text{for}~~-\tilde{b}\leq y < 0_- \\
\tilde{b} ~~&~~\text{for}~~~y=0_+.
\end{cases}
\label{tDelta_y-f<1/3}
\end{align}
Note the $y$ dependence of $\bar{\delta}_y$ for $0<f < 1/3$ is same as $\bar{\delta}_y$ for $1/3\leq f < 1/2$ given in Eq.~\eqref{eq:delta_tilde_bar} with $\tilde{x}_f=0$. Hence, the analysis of the SPE \eqref{eq:se_fless_edge_kmone_scaled} is almost same as for the $1/3<f<1/2$ case with $\tilde{x}_f=0$. Hence the calculations for $T_3$, $T_4$ and $T_5$ in the range $\tilde{b}\leq y\leq -\tilde{x}_f$ for the previous case ($1/3\leq f< 1/2$) would go through and the expressions of these terms for $0<f<1/3$ can be easily obtained by putting $\tilde{x}_f=0$ in the expressions of these integrals.  Hence, for  $0<f<1/3$ the SPE \eqref{eq:se_fless_edge_kmone_scaled} gets simplified to 
\beq
(1- \tilde{\varrho}_{\rm edge})\frac{y^2}{2} \approx \tilde{\mu}_2,
\eeq
which once again implies $\tilde{\varrho}_{\text{edge}} = 1$, and   $\tilde{\mu}_2 \approx 0$. The value for $\tilde{b}$ can be determined from the normalisation $\int_{-\tilde{b}}^{0_-}\tilde{\varrho}_f(y)dy=f$ and it yields $\tilde{b} \approx f$ as in Eq.~\eqref{eq:solution_finbetween}. One can follow the same procedure to solve the saddle point equation on the right part and arrive at the same solution for $\tilde{\varrho}_f(y)$.

\section{Crossover in the density profiles in Riesz gas models for other values of $k$}
\label{app:den-cross-other-k}
\begin{figure}[t]
	\centering		
	\includegraphics[width=1.0\textwidth]{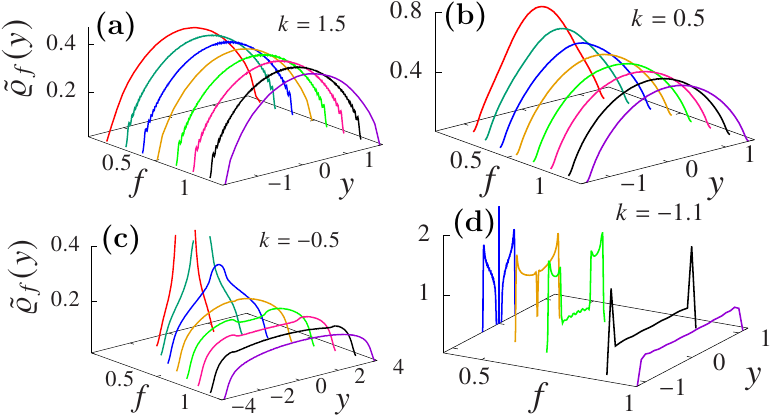}		
	\caption{This figure demonstrates how the density profiles change from one shape to another as $f$ is reduced from $1$ to $0$ for four representative values of $k$: (a) $k=1.5$, (b) $k=0.5$, (c) $k=-0.5$ and (d)$k=-1.1$, chosen from the following ranges $k>1$, $0<k<1$, $-1<k<0$ and $-2<k<-1$. For each case except $k=-1.1$, we have plotted  scaled density profiles $\varrho_f(y)$ for 8 values of $f$: $f=n/8$ with $n=8,7,...,1$ (from violet to red). To avoid congestion, for $k=-1.1$ we show density profiles only for five values of $f$ corresponding to $n=8,7,5,4$ and $3$ (from violet to blue). For $k=1.5$ there is neither a change in the support nor in the shape. For $k=0.5$, with decreasing $f$ the repulsion among the particles reduces which causes them to gather near the  centre of the trap. Consequently, the profile get squeezed  from dome shape at $f=1$ to a hill shape at $f \to 0$. For $k=-0.5$, the equilibrium density profile changes drastically as compared to the previous two cases. The profile shows a crossover from a dome-shaped profile at $f \to 1$ limit to a bell-shaped profile at $f \to 0$. Near $f=0$, the peaks of the bell-shaped profiles on the $z$-axis are cut in order to display the features of the density profiles for $f$ close to $1$. For  $k=-1.1$, the density profile changes from a `U' shaped profile at $f=1$ to a bell-shaped profile at $f=0$ passing through different other interesting shapes at intermediate values of $f$.} 
	\label{fig:density_crossover}
\end{figure}
For other values of $k>-2$ in Eq.~\eqref{gen_model}, we observe similar crossover in the density profile with respect to $f$ as observed for the $1$dOCP and log-gas models. In this section, we briefly demonstrate such  crossover. Based on the behavior of the density profiles due to the variation in $f$, one can categorise them into three $k$ regimes. \\
\noindent
(i) $k > 1$: In this case, we find that for any fixed $f\in (0,1]$ the density profile for large $N$ remains same as in the ATAC ($f=1$) case. 
The density profile posses the same scaling form $\varrho_N(x)=(1/N^{\alpha_k}) \tilde{\varrho}_f(x/N^{\alpha_k})$ with $\alpha_k=k/(k+2)$ 
and does not depend on $f$. This happens because for $k>1$, the free energy functional $\Sigma[\rho_N(x)]$ (in leading order in $N$) is local as the dominant contribution comes from the self energy contribution. In Fig.~\ref{fig:density_crossover}(a), we plot the scaled density profile for this case that are obtained numerically.  \\
\noindent
(ii) $0<k<1$: In this range of models,  the ATAC case ($f=1$) was studied in Ref.~\cite{sanaa2019} and the SR case ($f=0$) was studied in Ref.~\cite{Avanish_PRE_2020}. In these studies, it was observed that in both the limits of $f$, the equilibrium density profile is finitely supported and possesses the same scaling form $\varrho_N(x)=(1/N^{\alpha_k}) \tilde{\varrho}_f(x/N^{\alpha_k})$ as in the $k>1$ case. However, in this case the exponent $\alpha_k$ takes different values for $f=1$ and $f=0$ \cite{sanaa2019,Avanish_PRE_2020}. Consequently, the shape of the density profile differs. We numerically observe from Fig.~\ref{fig:density_crossover}(b) that with decreasing $f$ from one, the density profile seems to smoothly change from the ATAC profile to the SR profile.\\
\noindent
(iii) $-2<k<0$: Unlike the previous cases, in this range of $k$, we observe that the density profile undergoes drastic change of shape as $f$ is reduced from $1$ to $0$. Crossover in the density profiles in this case is demonstrated in Fig.~\ref{fig:density_crossover}(c) and Fig.~\ref{fig:density_crossover}(d).  For $-1 < k \le 0$, the structural shift in the density profile takes place through a transition from a dome-shaped profile at $f=1$ to a bell-shaped profile in the $f \to 0$ limit (see Fig.~\ref{fig:density_crossover}(c)). The particular case of   log-gas ($k \to 0$) falls in this regime which has been studied in detail in Sec.~\ref{crossover_genkgt0}. For  $-2 < k < -1$, the crossover in density profile, shown in Fig.~\ref{fig:density_crossover}(d), is qualitatively similar to the $1$dOCP ($k=-1$) case discussed in Sec.~\ref{density_crossover} of the main text. In the ATAC limit ($f=1$), the density profile has a shape of `U' with integrable divergences at the edges. As $f$ is decreased from $1$, two narrow density profiles with density values higher than that of the central part appear at the edges. The edge parts approach towards each other with  decreasing $f$ further until they touch each other at the centre of the trap at  $f=1/2$. When $f$ is decreased further, a fraction of particles gets accumulated at the centre of the trap and the rest of the fraction stay pushed away creating two holes (regions deprived of particles) symmetrically placed on both sides of the central peak.

\end{document}